\documentclass[journal]{IEEEtran}
\usepackage{amssymb}

\usepackage{cite}

\ifCLASSINFOpdf
   \usepackage[pdftex]{graphicx}
\else
\fi
\usepackage{amsmath}
\begin{document}
%
\title{Variation of a Signal in Schwarzschild Spacetime}
%
%
%

\author{Huan~Liu,
        Xiang-Gen~Xia,~\IEEEmembership{Fellow,~IEEE,}
        and~Ran~Tao,~\IEEEmembership{Senior Member,~IEEE}
        
\thanks{H. Liu, X.-G. Xia, and R. Tao are with the School of Information and Electronics, Beijing Institute of Technology, Beijing, 100081, P. R. China. E-mail: (liuhuandemengxiang@163.com; xianggen@udel.edu; rantao@bit.edu.cn).}

\thanks{X.-G. Xia is also with the Department
of Electrical and Computer Engineering, University of Delaware, Newark,
DE 19716, USA.}
\thanks{This work was supported in part by the National Natural
Science Foundation of China under Grant 61421001, Grant
61331021 and Grant U1833203.}
}

%
%

\markboth{}%
{Shell \MakeLowercase{\textit{et al.}}: Variation of a Signal in Schwarzschild Spacetime}
%



\maketitle

\begin{abstract}
In this paper, the variation of a signal in Schwarzschild spacetime is studied and a general equation for frequency shift parameter (FSP) is presented. It shows that FSP depends on the gravitationally-modified Doppler effects and the gravitational effects of observers. In addition, rates of time of a transmitter and a receiver may be different. When FSP is a function of the time of a receiver, FSP contributed by gravitational effect (GFSP) or gravitationally-modified Doppler effect (GMDFSP) may lead a bandlimited signal to a non-bandlimited signal. Based on the equation, FSP as a function of the time of a receiver is calculated for three scenarios: a) a spaceship moves away from a star with a constant velocity communicating with a transmitter at a fixed position; b) a spaceship moves around a star with different conic trajectories communicating with a transmitter at a fixed position; c)  a signal is transmitted at a fixed position in a star system to a receiver moving with an elliptical trajectory in another star system.  The studied stars are sun-like star, white dwarf and neutron star, and some numerical examples are presented to illustrate the theory.
\end{abstract}


%
\IEEEpeerreviewmaketitle

\section{Introduction}
%
%
%
%
\IEEEPARstart{D}{ue}
to the development of space technologies, human beings have  sent spacecrafts to  explore the space far away from Earth. After exploring the planets in the solar system, Voyager 1 of Voyager Interstellar Mission of NASA escaped the solar system and entered the interstellar medium on August 25, 2012 and is on its journey to explore the outer space~\cite{Voyager12018}, and Voyager 2 was at a distance of 119 AU from the sun on November 02, 2018 and is going to achieve the escape velocity to leave the solar system and explore interstellar space~\cite{Voyager22018}. The success of a series of Chang'E missions inspired China to initial its first Mars Program on January 11, 2016~\cite{Zhao2018Network}. Focusing on deep space communications (but without considering the gravitational effect), many aspects have been discussed recently, including communication network design~\cite{Zhao2018Network,Wu2018Design,Wan2018Solar}, channel models~\cite{Pan2018Review}, channel coding~\cite{Wu2018Novel}, laser communications~\cite{Wu2018Overview}. It is not hard to imagine that mankind will be able to take high-speed  spaceship to explore even much deeper space in the near future. Therefore, the study of deep space communications becomes greatly important.

After Einstein field equations of general relativity were proposed~\cite{Einstein1915Die}, Schwarzschild proposed the first solution, called Schwarzschild metric\cite{Schwarzschild1916b1,Schwarzschild1916b2}, which describes the curved spacetime around a static space object with spherical symmetry. Schwarzschild metric predicts that when an electromagnetic wave with a certain frequency at a position propagates to another position, the frequency changes as long as the gravities at the two positions are unequal. If the gravity at the latter is bigger (smaller) than that at the former, the rate of time of  the latter is smaller (bigger) and the frequency blue (red) shifts. These phenomena have already been proven by a series of experiments~\cite{Pound1959Gravitational,Pound1965Effect,Snider2001New,Turner1964New}, and have been taken into account for satellite tracking~\cite{harkins1979relativistic}, Global Positioning System (GPS)~\cite{Love2002GPS} and X-ray pulsar-based navigation by weak field approximation~\cite{hanson1991principles,sheikh2005use,sheikh2006spacecraft}.

For wireless communications near the earth, Doppler effect plays an important role and communication signal frequency varies according to	the relative velocity. 
Gravitational effect has merely been taken into account as the gravitation of Earth is not great enough to noticeably affect the spectrum of a signal. However, For wireless communications in the deep space, we will see that Doppler effect is still important (also see \cite{oberg2004titan}) but should be modified, and gravitational effect becomes much more important since it may involve not only  Earth but also the sun, white dwarf, neutron star or other space objects with large masses. Instead of depending on the relative velocity, the variation of frequency by gravitational effect depends on the positions of a transmitter and a receiver.    

In this paper, gravitational effect on a signal is studied based on Schwarzschild metric of general relativity. The studied stars  are sun-like star, white dwarf and neutron star and three different scenarios are considered: a) a spaceship moves away from a star with a constant velocity communicating with a transmitter at a fixed position; b) a spaceship moves around a star with different conic trajectories communicating with a transmitter at a fixed position; c)  a signal is transmitted from a star system to another star system where the receiver is in an elliptical trajectory. The studies in this paper may help to better understand the signal models for deep space communications.

This paper is organized as follows. In Section~\ref{section2}, a general equation for frequency shift parameter (FSP) in Schwarzshild  spacetime is presented. In Section~\ref{section3}, based on the general equation developed in Section~\ref{section2}, the behaviors of a signal traveling in Schwarzschild spacetime are studied for the three scenarios a), b) and c) as described above. In Section~\ref{section4}, numerical examples for the three scenarios are presented. 
 
Some notations  used in this paper are presented as follows~\cite{MisnerGravitation1973,WaldGeneral1984}:

$T$: time coordinate or Newtonian time or time of a static observer at the infinity in Schwarzschild spacetime or time of a static observer in Minkowski spacetime.

$t$: time of a static observer at $r$ in Schwarzschild spacetime.

$\tau$: time of a moving observer at $r$ in Schwarzschild spacetime.

$\frac{\partial} {\partial T}$:  a coordinate basis vector of  partial derivative operator along direction $T$.

a boldface English letter is a  vector represented  by a set of coordinate basis vectors. For example, ${\bf V}=V_0\frac{\partial}{\partial T}+V_1\frac{\partial}{\partial r}+V_2\frac{\partial}{\partial \theta}+V_3\frac{\partial}{\partial \varphi}$, where $V_m$ is real-valued and the $m$th component of vector ${\bf V}$ ($m$=0, time coordinate component; $m$=1, 2, 3, space coordinate components). 

${\bf d}T$:  the dual vector of $\frac{\partial} {\partial T}$ along direction $T$, which is a linear operator on a vector space, i.e., ${\bf d}T (\frac{\partial} {\partial T})=\frac{\partial T} {\partial T}=1$, and ${\bf d}T({\bf V})=V_0\frac{\partial T}{\partial T}+V_1\frac{\partial T}{\partial r}+V_2\frac{\partial T}{\partial\theta}+V_3\frac{\partial T}{\partial \varphi}=V_0+V_1\cdot 0+V_2 \cdot 0+V_3\cdot 0=V_0$. 

${\bf d}T\otimes {\bf d}T$:  a coordinate basis tensor which is the tensor product of ${\bf d}T$ and ${\bf d}T$ and a bi-linear operator on a vector space, i.e., ${\bf d}T\otimes {\bf d}T\left(\frac{\partial}{\partial T},\frac{\partial}{\partial T}\right)={\bf d}T(\frac{\partial}{\partial T})\cdot{\bf d}T(\frac{\partial}{\partial T})=1\cdot1=1$, and ${\bf d}T\otimes{\bf d}T({\bf V},{\bf W})={\bf d}T({\bf V})\cdot {\bf d}T({\bf W})=V_0W_0$.

${\bf g}$: Schwarzshild metric in Schwarzshild spacetime which is a bi-linear operator on a vector space and can be represented by a set of coordinate basis tensors (see (\ref{eq1}) in the following).
 
$g_{mn}$: the $(m,n)$th component of   metric   ${\bf g}$.

${\bf g}\left({\bf V}, {\bf W}\right)${\rm =}$\sum_{m,n}{g_{mn}V_{m}W_{n}}$.

${\bf g}\left({\bf V}, {\bf V}\right)$=$\sum_{m,n}{g_{mn}V_{m}V_{n}}$. If it is positive, ${\bf V}$ is a space-like vector; if it  is negative, ${\bf V}$ is a time-like vector; if it is zero, ${\bf V}$ is a light-like vector and vice versa. It should be noted that  a time-like vector may have space coordinate components and a space-like vector may have a time coordinate component.

${\bf g}\left({\bf V}, {\bf V}\right)=1$:  ${\bf V}$ is a unit space-like vector.

${\bf g}\left({\bf V},{\bf V}\right)=-1$:   ${\bf V}$ is a unit time-like vector.

${\bf h}$: Minkowski metric in Minkowski spacetime which is a bi-linear operator on a vector space and can be represented by a set of coordinate basis tensors (see (\ref{eq15}) in the following).

\section{Frequency change in Schwarzschild spacetime}{\label{section2}}
In Newtonian mechanics, the world is described by three dimensional space and one dimensional time with a fixed rate of time for all observers. Time and space are treated separately.   A signal is a function of this time and its spectrum changes according to Doppler equation. In special relativity, the world is described by four dimensional Minkowski spacetime where a moving observer's own time has relation with the observer's
velocity and thus time and space are mixed together. Rates of time are equal for all static observers but are different from moving observers' rates of time. A signal is a function of its observer's own time determined by the observer' velocity and its spectrum changes according to the special relativity version of Doppler equation. In general relativity, the world is also described by four dimensional spacetime where time and space are mixed together as well but  further take into account of  gravitational effect. Rates of time may be not equal for static observers at different positions. A signal is  a function of its observer's own time determined by the observer's velocity and position. Doppler equation should be further modified as we shall see later, and gravitational effect should be considered in calculating the change of its spectrum. Therefore, it is more complex and we will derive them next.

Schwarzschild spacetime is one solution of Einstein field equations in general relativity and describes space and time around a static object with spherical symmetry. Schwarzschild spacetime is characterized by Schwarzschild metric as follows~\cite{Schwarzschild1916b1,Schwarzschild1916b2,MisnerGravitation1973,WaldGeneral1984}:
\begin{equation}{\label{eq1}}
\begin{small}
\begin{aligned}
{\bf g}\triangleq& g_{00}{\bf d}T\otimes{\bf d}T+g_{11}{\bf d}r\otimes{\bf d}r+g_{22}{\bf d}\theta\otimes{\bf d}\theta+g_{33}{\bf d}\varphi\otimes{\bf d}\varphi\\
=&-c^{2}\left(1-\frac{2GM}{c^{2}r}\right){\bf d}T\otimes{\bf d}T+\left(1-\frac{2GM}{c^{2}r}\right)^{-1}{\bf d}r\otimes{\bf d}r\\
&+r^2\left({\bf d}\theta\otimes{\bf d}\theta+\sin^2\theta{\bf d}\varphi\otimes{\bf d}\varphi\right),
\end{aligned}
\end{small}
\end{equation}
where $G$ is the gravitational constant, $c$ represents the speed of light, $M$ denotes the mass of the static object with spherical symmetry (sun, white dwarf and neutron star are regarded as static and spherically symmetric even though they may  be neither static nor perfectly spherically symmetric) and $ \{r,\theta ,\phi\}$ are the spherical coordinates of the reference frame whose origin is the center of the object. For simplicity, we also denote ${\bf g}=(g_{mn})_{0\leq m, n\leq 3}$ and $\alpha\triangleq 2GM/c^2$. In general, $\alpha$ is  smaller than $r$, otherwise it is a black hole that is not discussed in this paper. Therefore, $g_{00}<0$, $g_{nn}>0$ for $n=1,2,3$ and $g_{mn}=0$ for $m\neq n$, which is the reason why the “squared length” (weighted with weights $g_{mn}$) ${\bf g}({\bf V},{\bf V})$ of a vector ${\bf V}$ can be zero and even negative.

From (\ref{eq1}), the unit vectors of four coordinates in Schwarzschild spacetime can be obtained. It should be noted that a unit time-like  vector ${\bf V}$ satisfies ${\bf g}\left({\bf V},{\bf V}\right)=-1$ and a unit space-like vector ${\bf V}$ satisfies ${\bf g}\left({\bf V}, {\bf V}\right)=1$. One unit time coordinate  vector and three unit space coordinate vectors are, respectively, presented as follows:
\begin{equation} {\label{eq2}}
\begin{aligned}
{\bf Z}=\frac{1}{c}\frac{\partial}{\partial t}&=\frac{1}{c}(1-\alpha/r)^{-\frac{1}{2}}\frac{\partial}{\partial T},\\ 
(1-\alpha/r)^{\frac{1}{2}}\frac{\partial}{\partial r},&\,\,\, 
r^{-1}\frac{\partial}{\partial\theta}, \,\,\,
r^{-1}\sin^{-1}\theta\frac{\partial}{\partial\varphi}.
\end{aligned}
\end{equation}
Note that, $T$ is not only time coordinate but also represents the time at the infinity  where there is no gravitation (or Newtonian time that does not take into account of the influence of gravitation). The vector ${\bf Z}$ is just the unit vector of a static observer's time direction at $r$ and the observer measures the time based on it, and $t$ is the time of the observer. The dual vector of ${\bf Z}$ is 
\begin{equation}{\label{eq3}}
\begin{small}
\begin{aligned}
&c{\bf d}t=c(1-\alpha/r)^{\frac{1}{2}}{\bf d}T,\\
\end{aligned}
\end{small}
\end{equation}
and from the dual vector, according to \cite{MisnerGravitation1973}, we can obtain:
\begin{equation}{\label{eq4}}
\begin{small}
\begin{aligned}
&dt=(1-\alpha/r)^{\frac{1}{2}}dT,\\
\end{aligned}
\end{small}
\end{equation}
where the relation between $dt$ and $dT$ is dependent of $r$. It shows that if the rate of time at the infinity (or the rate of Newtonian time) is 1, the rate of time at $r$ is relatively $(1-\alpha/r)^{\frac{1}{2}}$. Therefore,  rates of time are different at different $r$, and the bigger $r$ is, the bigger rate of time is.  

It is known that an electromagnetic wave is described by two parts separately: an angular frequency $\omega$ and a spatial wave vector ${\bf k}$. In Schwarzschild spacetime, the two parts are combined together into a four dimensional wave vector~\cite{WaldGeneral1984}:
\begin{equation}{\label{eq5}}
\begin{small}
{\bf K}=\omega {\bf Z}+c{\bf k},
\end{small}
\end{equation}
where  ${\bf k}$ is a space-like vector and its time coordinate component is zero. In the geometrical optics approximation, ${\bf K}$ is a light-like vector \cite{WaldGeneral1984}, i.e, ${\bf g}({\bf K}, {\bf K})=0$. {\bf Z } is a unit time-like vector without space coordinate component, i.e., ${\bf g}({\bf Z}, {\bf Z})=-1$ and ${\bf k}$ is a space-like vector without time coordinate component, i.e.,  ${\bf g}({\bf Z}, {\bf k})=0$. Thus, it is not hard to obtain:
\begin{equation}{\label{eq6}}
\begin{small}
{\bf g}\left({\bf k}, {\bf k}\right)=\omega^{2}/c^{2}.
\end{small}
\end{equation} 

When two static observers are at different positions, say $\{r_{1},r_{2}\}$,  and  the first observer at $r_1$ transmits an electromagnetic wave with angular frequency  $\omega_1$ to the second observer, frequency $\omega_2$ of the received wave at $r_2$ obeys the following equation \cite{WaldGeneral1984}:
\begin{equation}{\label{eq7}}
\begin{small}
\begin{aligned}
\omega_{2}
&=\omega_1\left(\frac{1-\alpha/r_1}{1-\alpha/r_2}\right)^{\frac{1}{2}}.
\end{aligned}
\end{small}
\end{equation}
It can be seen from (\ref{eq5}) that angular frequency $\omega$ is actually the projection value after projecting ${\bf K}$ along ${\bf Z}$, i.e., the unit vector of a static observer's time direction. Therefore, if the electromagnetic wave vector transmitted by a static observer at $r_1$ is 
\begin{equation}{\label{eq8}}
\begin{small}
 {\bf K}_{r_1}=\omega_1{\bf Z}_{r_1}+c{\bf k}_{r_1},
 \end{small}
\end{equation}
then, the electromagnetic wave vector received by a static observer at  $r_2$ is  
\begin{equation}{\label{eq9}}
\begin{small}
{\bf K}_{r_2}=\omega_2{\bf Z}_{r_2}+c{\bf k}_{r_2},
\end{small}
\end{equation}
where ${\bf Z}_{r_i}$ and ${\bf k}_{r_i}$  are unit time coordinate vector and spatial wave vector for the static  observer at  $r_i$ for $i=1,2$, respectively. In addition, from (\ref{eq4}) and (\ref{eq7}), it can be seen: 
\begin{equation}{\label{eq10}}
\begin{small}
\begin{aligned}
\omega_{2}/\omega_{1}=dt_1/dt_2,
\end{aligned}
\end{small}
\end{equation}
where $t_{i}$ is the time of the static observer at  $r_{i}$ for $i=1,2$. Therefore, angular frequency $\omega$ is inversely proportional to rate of time if two observers are static. Obviously, if $r_2>r_1$, the frequency red shifts and  if $r_1<r_2$, the frequency blue shifts. 
\begin{figure}[t!]
\centering
\includegraphics[scale=0.45]{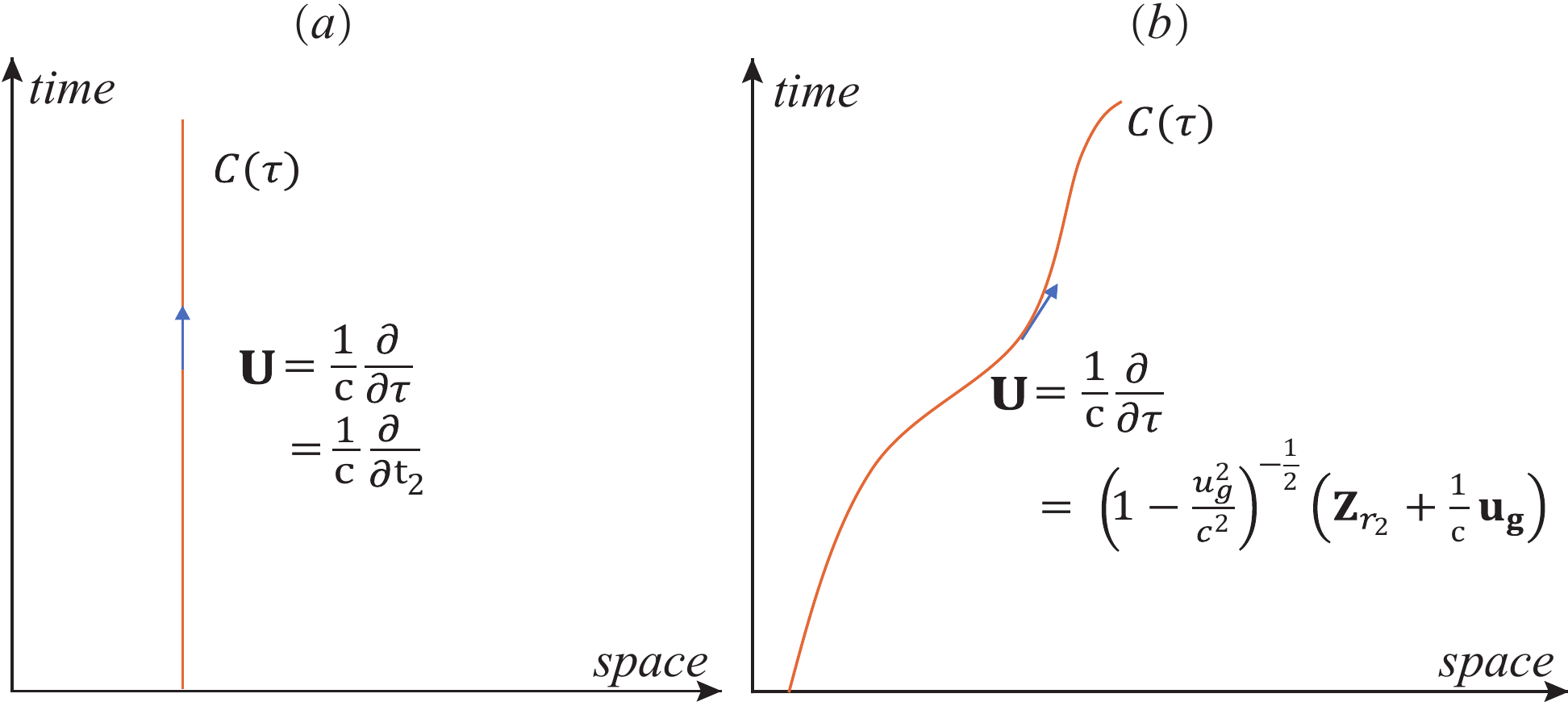}
\caption{World lines of $(a)$ a static observer and $(b)$ a moving observer in Schwarzschild spacetime.}
\label{ppt}
\end{figure}

Now, consider that a new observer at $r_2$ is not static but with four dimensional velocity (also called 4-velocity~\cite{WaldGeneral1984}):
\begin{equation}{\label{eq11}}
\begin{small}
\begin{aligned}
{\bf U}&=\frac{1}{c}\frac{\partial}{\partial \tau}\\
&=\frac{dt_2}{d\tau}{\bf Z}_{r_2}+\frac{1}{c}\bigg(\frac{dr_2}{d\tau}\frac{\partial}{\partial r_2}+\frac{d\theta}{d\tau}\frac{\partial}{\partial \theta}+\frac{d\varphi}{d\tau}\frac{\partial}{\partial \varphi}\bigg)\\
&=\frac{dt_2}{d\tau}{\bf Z}_{r_2}+\frac{1}{c}\bigg(\frac{dr_2/dt_2}{d\tau/dt_2}\frac{\partial}{\partial r_2}+\frac{d\theta/dt_2}{d\tau/dt_2}\frac{\partial}{\partial \theta}+\frac{d\varphi/d\tau}{d\tau/dt_2}\frac{\partial}{\partial \varphi}\bigg)\\
&=\frac{dt_2}{d\tau}\bigg[{\bf Z}_{r_2}+\frac{1}{c}{\bf u_g}\bigg],\\
\end{aligned}
\end{small}
\end{equation}
where 
\begin{equation}{\label{eq12}}
\begin{small}
\begin{aligned}
{\bf u_g}&\triangleq\frac{dr_2}{dt_2}\frac{\partial}{\partial r_2}+\frac{d\theta}{dt_2}\frac{\partial}{\partial \theta}+\frac{d\varphi}{dt_2}\frac{\partial}{\partial \varphi}\\
&=(1-\alpha/r_2)^{-\frac{1}{2}}\bigg(\frac{dr_2}{dT}\frac{\partial}{\partial r_2}+\frac{d\theta}{dT}\frac{\partial}{\partial \theta}+\frac{d\varphi}{dT}\frac{\partial}{\partial \varphi}\bigg)
\end{aligned}
\end{small}
\end{equation}
is the spatial velocity of the new observer in Schwarzschild spacetime and is a space-like vector without time coordinate component. The second equality in (\ref{eq12}) is according to (\ref{eq4}) with $r=r_2$ and $t=t_2$. Let $u_g=\left[{\bf g}({\bf u_g}, {\bf u_g})\right]^{\frac{1}{2}}$, which is the magnitude of the spatial velocity ${\bf u_g}$ at $r_2$ in Schwarzschild spacetime~\cite{WaldGeneral1984}. Then, 
\begin{equation}{\label{eq13}}
\begin{small}
\begin{aligned}
u_g
=&\big[(1-\alpha/r_2)^{-1}(dr_2/dt_2)^2+r_2^2(d\theta/dt_2)^2\\
&+r_2^2\sin^2\theta(d\varphi/dt_2)^2\big]^{\frac{1}{2}}\\
=&(1-\alpha/r_2)^{-\frac{1}{2}}\big[(1-\alpha/r_2)^{-1}(dr_2/dT)^2+r_2^2(d\theta/dT)^2\\
&+r_2^2\sin^2\theta(d\varphi/dT)^2\big]^{\frac{1}{2}},
\end{aligned}
\end{small}
\end{equation}
where  the term $(1-\alpha/r_2)^{-1}$ in  the first equality is $g_{11}$  from $\bf g$.

It should be noted that ${\bf U}$ is the unit  vector of the new observer's time direction. The new observer measures the time based on ${\bf U}$ and $\tau$ is the time of the new observer.  To be more explicit, two examples are illustrated in Fig.~\ref{ppt}. The coordinates in Fig.~\ref{ppt} are combinations of three dimensional space and one dimensional time into four dimensional spacetime  (only two dimensions are depicted for simplicity). A curve $C(\tau)$ in the coordinates is time vs. space and called an observer's world line which is parametrized by $\tau$ or characterized by tangent vector $\frac{1}{c}\frac{\partial}{\partial\tau}$. A time-like curve means that every tangent vector of the curve is a unit time-like vector~\cite{MisnerGravitation1973}. Since every observer should not be faster than light, every world line of an observer is a time-like curve, $\tau$ is precisely the time of the observer along the curve and $\frac{1}{c}\frac{\partial}{\partial\tau}$ is like a ruler for the observer to measure the time.  In sub-figure (a), an observer at $r_2$ is static,  the tangent vector of every position on the curve is  $\frac{1}{c}\frac{\partial}{\partial t_2}$, and thus the observer's time is $t_2$. In sub-figure (b), an observer is moving with different velocities at different positions and its tangent vector is changing all the time, but at every position of its world line, it measures the time along the curve using the tangent vector as a ruler. Since the 4-velocity ${\bf U}$ is a unit time-like vector, i.e., ${\bf g}({\bf U},{\bf U})=-1$,  ${\bf Z}_{r_2}$ is a unit time-like vector without space coordinate component, and ${\bf u_g}$ is a space-like vector without time coordinate component, we have ${\bf g}({\bf Z}_{r_2},{\bf Z}_{r_2})=-1$ and ${\bf g}({\bf Z}_{r_2},{\bf u_g})=0$. Then, it is not hard to obtain:
\begin{equation}{\label{eq14}}
\begin{small}
dt_2/d\tau=(1-u_g^{2}/c^{2})^{-\frac{1}{2}}\triangleq\gamma_g.
\end{small}
\end{equation}

For the comparison purpose, in special relativity, the four dimensional spacetime is Minkowski spacetime. The metric characterizing the Minkowski spacetime is Minkowski metric (also called flat metric)~\cite{WaldGeneral1984}:
\begin{equation}{\label{eq15}}
\begin{small}
{\bf h}=-c^{2}{\bf d}T\otimes{\bf d}T+{\bf d}r\otimes{\bf d}r+r^2\left[{\bf d}\theta\otimes{\bf d}\theta+\sin^2{\theta}{\bf d}\varphi\otimes{\bf d}\varphi\right],
\end{small}
\end{equation}
and let ${\bf u_h}$ be the spatial velocity in Minkowski spacetime. Then,
\begin{equation}{\label{eq16}}
\begin{small}
{\bf u_h}=\frac{dr_2}{dT}\frac{\partial}{\partial r_2}+\frac{d\theta}{dT}\frac{\partial}{\partial \theta}+\frac{d\varphi}{dT}\frac{\partial}{\partial\varphi},
\end{small}
\end{equation}
and the magnitude $u_h$  of the spatial velocity ${\bf u_h}$ in Minkowski spacetime can be calculated:
\begin{equation}{\label{eq17}}
\begin{small}
\begin{aligned}
u_h&=\left[{\bf h}({\bf u_h}, {\bf u_h})\right]^{\frac{1}{2}}\\
&=\left[\left(dr_2/dT\right)^2+r_2^2\left(d\theta/dT\right)^2+r_2^2\sin^2\theta\left(d\varphi/dT\right)^2\right]^{\frac{1}{2}}.
\end{aligned}
\end{small}
\end{equation}
It can be seen from (\ref{eq12}) and (\ref{eq16}) that one difference of the spatial velocities in the two spacetimes is that Newtonian time $T$ (a static observer's time at $r_2$ in Minkowski spacetime) in  (\ref{eq16}) is just replaced by time $t_2$ (a static observer's time at $r_2$ in Schwarzschild spacetime) in (\ref{eq12}), though they share the same direction. Another difference is in the magnitudes of the spatial velocities in the two spacetimes. By comparing (\ref {eq13}) and (\ref{eq17}), there is one more coefficient $(1-\alpha/r_2)^{-1}$ in the first equality of  (\ref{eq13}) which is derived from $g_{11}$ in ${\bf g}$, in addition to the time difference of $T$ and $t_2$.

According to (\ref{eq14}), and (\ref{eq4}) with $r=r_2$ and $t=t_2$, we have 
\begin{equation}{\label{eq18}}
\begin{small}
\begin{aligned}
d\tau
&=(1-u_g^{2}/c^{2})^{\frac{1}{2}}(1-\alpha/r_2)^{\frac{1}{2}}dT.
\end{aligned}
\end{small}
\end{equation}
Since the frequency $\omega'_2$  at $r_2$ for the new observer is the component of its own time direction,  it is needed to project ${\bf K}_{r_2}$ along ${\bf U}$, i.e., the unit vector of the new observer's time direction, and the projection value is
\begin{equation}{\label{eq19}}
\begin{small}
\begin{aligned}
\omega'_{2}&=-{\bf g}\left({\bf U},{\bf K}_{r_2}\right)\\
&=-\gamma_g {\bf g}({\bf Z}_{r_2}+{\bf u_g}/c, \omega_2{\bf Z}_{r_2}+c{\bf k}_{r_2})\\
&=\gamma_g\omega_2-\gamma_g {\bf g}\left({\bf u_g},{\bf k}_{r_2}\right)\\
&=\gamma_g\omega_{2}-\gamma_g u_g\frac{\omega_{2}}{c}\cos{\psi}\\
&=\omega_{1}\gamma_g\left(1-\frac{u_g}{c}\cos{\psi}\right)\left(\frac{1-\alpha/r_{1}}{1-\alpha/r_{2}}\right)^{\frac{1}{2}},
\end{aligned}
\end{small}
\end{equation}
where $\omega_2$ is the frequency for a static observer at $r_2$ who shares the same ${\bf K}_{r_2}$ with the moving observer at $r_2$ but has different time direction, and $\cos\psi={\bf g}\left({\bf u_g},{\bf k}_{r_2}\right)/\sqrt{{\bf g}\left({\bf k}_{r_2},{\bf k}_{r_2}\right){\bf g}\left({\bf u_g},{\bf u_g}\right)}$. The second equality in  (\ref{eq19}) is  according to (\ref{eq9}), (\ref{eq11}) and (\ref{eq14}), the second last equality is according to  $(\ref{eq6})$ with ${\bf k}={\bf k}_{r_2}$, and the last equality is according to  $(\ref{eq7})$.  We denote
\begin{equation}{\label{eq20}}
\begin{small}
\beta\triangleq\gamma_g\left(1-\frac{u_g}{c}\cos{\psi}\right)\left(\frac{1-\alpha/r_{1}}{1-\alpha/r_{2}}\right)^{\frac{1}{2}},
\end{small}
\end{equation}
and $\beta$ is called {\em frequency shift parameter} (FSP)
and is independent of angular frequency $\omega_1$. Thus, the frequency $\omega'_2$ at $r_2$ of the moving observer in terms of the signal frequency $\omega_1$ of the transmitter is
\begin{equation}{\label{neweq19}}
\omega'_2=\beta\omega_1.
\end{equation}
It is not hard to see that FSP $\beta$ in  (\ref{eq20}) can be separated into two parts. 
The first part is 
\begin{equation}{\label{eq21}}
\begin{small}
\beta_1\triangleq\gamma_g\left(1-\frac{u_g}{c}\cos{\psi}\right),
\end{small}
\end{equation}
which is similar to the special relativity version of the Doppler equation \cite{WaldGeneral1984},
\begin{equation}{\label{eq22}}
\begin{small}
\beta'_1=\gamma_h\left(1-\frac{u_h}{c}\cos\psi'\right),
\end{small}
\end{equation}
where $\gamma_h$ is similar to $\gamma_g$ in  (\ref{eq14}) but with $u_g$ replaced by $u_h$, and $\psi'$ is the intersection angle between ${\bf u}_h$ and ${\bf k}_{r_2}$. When $r_1, r_2\rightarrow\infty$, (\ref{eq21}) and (\ref{eq22}) are equivalent. Furthermore, if $u_h\ll c$, then $\gamma_h\approx 1$ and we have 
\begin{equation}{\label{eq23}}
\begin{small}
\beta'_1 =(1-\frac{u_h}{c}\cos{\psi'}),
\end{small}
\end{equation}
which is the conventional version of the Doppler equation. Therefore,  (\ref{eq21}) is just a gravitationally-modified Doppler equation and $\beta_1$ is called {\em gravitationally-modified Doppler frequency shift parameter} (GMDFSP). 

The second part is purely derived from gravitational effect (or called Einstein effect in \cite{moller1972relativity}), and is called {\em gravitational frequency shift parameter} (GFSP): 
\begin{equation}{\label{eq24}}
\begin{small}
\beta_2\triangleq\left(\frac{1-\alpha/r_{1}}{1-\alpha/r_{2}}\right)^{\frac{1}{2}}.
\end{small}
\end{equation}
Thus, FSP can be rewritten as
\begin{equation}{\label{eq25}}
\begin{small}
\beta=\beta_1\beta_2.
\end{small}
\end{equation}

It is clear that when $\beta$ is much different from 1, $\omega_2'$ and $\omega_1$ in  (\ref{neweq19}) are much different, i.e., the signal frequency of a moving observer at $r_2$ and the signal frequency of a transmitter at $r_1$ are much different. This means that there is a large frequency shift. Since $\beta$ is comprised of $\beta_1$ and $\beta_2$, the contributions from the two parts can be considered separately. If the difference of $\beta_1$ and 1 (or $\beta_2$ and 1) is large or small, gravitationally-modified Doppler effect (or gravitational effect) is significant or negligible.

When an observer at $r_2$ is static, (\ref{eq19}) comes back to  (\ref{eq7}), since $u_g=0$ and $\gamma_g=1$, and the FSP $\beta$ in  (\ref{eq20}) is equal to $\beta_2$ and is constant.  Also, when $u_g$ or $\psi$ in  (\ref{eq21}) changes over the time $\tau$, $\beta_1$ is a function of time $\tau$, and when $r_1$ or $r_2$ in  (\ref{eq24}) changes over the time $\tau$, $\beta_2$ is also a function of time $\tau$. As a result, $\beta$ in  (\ref{eq20}) and the frequency $\omega_2'$ in  (\ref{neweq19}) are functions of time $\tau$. In addition, $0<\beta_1(\tau)<\infty$ since $u_g<c$, and $0<\beta_2(\tau)<\infty$ since $\alpha<r_1$ and  $r_2$, in general. Therefore, $0<\beta(\tau)<\infty$. 

Equation (\ref{eq20}) is a basic equation for calculating FSP $\beta$. It depends on parameters $r_1$, $u_g$, $r_2$ and $\cos\psi$, which are related to the trajectory of a receiver. Some specific calculations of FSP $\beta$ will be presented for different receiver trajectories  in Section \ref{section3}.
\section{ A signal in Schwarzschild spacetime}{\label{section3}}
After the basic principle has been established above for the FSP when an electromagnetic wave is propagating through Schwarzschild spacetime, the variation for a signal transmitted in Schwarzschild spacetime in a particular scenario is not hard to see. In this section, to be consistent with Section~\ref{section2}, we denote the time of a transmitter at a fixed position $r_1$, the time of a moving receiver at $r_2$ and the time at infinity  as $t_1$, $\tau$ and $T$, respectively. Without considering noise, when a single-frequency complex exponential signal $e^{j\omega_1 t_1}$ is transmitted at $r_1$ to a spaceship at $r_2$, the received signal by the spaceship will be different,  that is:\\
\begin{equation}{\label{eq26}}
\begin{small}
e^{j\omega_{1}t_1}\rightarrow e^{j\omega_1B(\tau)},
\end{small}
\end{equation}
where $B(\tau)=\int_{0}^{\tau}{\beta(\tau')}d\tau'$, and $\beta(\tau')$ is the FSP in  (\ref{eq20}). If $\beta$ is constant, $B(\tau)=\beta\tau$. A periodical signal $f_1(t_1)$ with fundamental angular frequency $\omega$ can be expanded as Fourier series:
\begin{equation}{\label{eq28}}
\begin{small}
 f_1(t_1)=\sum^{\infty}_{k=-\infty}{a_{k}e^{jk\omega t_1}},
 \end{small}
\end{equation}
and
\begin{equation}{\label{eq29}}
\begin{small}
a_k=\frac{\omega}{2\pi}\int_{-\pi/\omega}^{\pi/\omega}{f(t_1)e^{-jk\omega t_1}dt_1}.
\end{small}
\end{equation}
Then, the received signal is
\begin{equation}{\label{eq30}}
\begin{small}
f_2(\tau)=\sum^{\infty}_{k=-\infty}{a_{k}e^{jk\omega B(\tau)}}.
\end{small}
\end{equation}

For a non-periodic signal, its fundamental angular frequency can be regarded as infinitesimal $\Delta\omega$. Then, the transmitted signal can be expanded as~\cite{OppenheimSignals1996}: 
\begin{equation}{\label{eq31}}
\begin{small}
f_1(t_1)=\sum^{\infty}_{k=-\infty}{a_{k}e^{jk\Delta\omega t_1}},
\end{small}
\end{equation}
and
\begin{equation}{\label{eq32}}
\begin{small}
a_k=\frac{\Delta\omega}{2\pi}\int_{-\pi/\Delta\omega}^{\pi/\Delta\omega}{f_1(t_1)e^{-jk\Delta\omega t_1}dt_1}.
\end{small}
\end{equation}
Then, the recieved signal is
\begin{equation}{\label{eq33}}
\begin{small}
\begin{aligned}
f_2(\tau)&=\sum^{\infty}_{k=-\infty}{a_{k}e^{jk\Delta\omega \tau}}\\
&=\sum^{\infty}_{k=-\infty}{\frac{\Delta\omega}{2\pi}\int_{-\pi/\Delta\omega}^{\pi/\Delta\omega}{f_1(t_1)e^{-jk\Delta\omega t_1}dt_1}e^{-jk\Delta\omega B(\tau)}}.
\end{aligned}
\end{small}
\end{equation}
As $\Delta\omega\rightarrow 0$, the received signal can be rewritten as
\begin{equation}{\label{eq34}}
\begin{small}
\begin{aligned}
f_2(\tau)&=\frac{1}{2\pi}\int_{-\infty}^{\infty}{\int_{-\infty}^{\infty}{f_1(t_1)e^{-j\omega t_1}dt_1}e^{j\beta\omega \tau}d\omega}\\
&=\frac{1}{2\pi}\int_{-\infty}^{\infty}{\hat{f}_1(\omega)e^{j\beta\omega \tau}d\omega}
=f_1(B(\tau)),
\end{aligned}
\end{small}
\end{equation}
where $\hat{f}_1(\omega)$ is the Fourier transform of $f_1(t_1)$.

In general, a communication signal is bandlimited and a bandlimited signal of bandwidth $W$ is
\begin{equation}{\label{eq35}}
\begin{small}
f_1(t_1)=\frac{1}{2\pi}\int_{-W}^{W}{\hat{f}_1(\omega)e^{j\omega t_1}d\omega}.
\end{small}
\end{equation}
If $\beta$ is a non-zero constant, according to (\ref{eq34}), the received signal is 
\begin{equation}{\label{eq36}}
\begin{small}
f_2(\tau)=\frac{1}{2\pi}\int_{-\beta W}^{\beta W}{\frac{1}{\beta}\hat{f}_1(\frac{\omega}{\beta})e^{j\omega \tau}d\omega}.
\end{small}
\end{equation}
Then,  the spectrum of the received signal is broadened when $\beta<1$, narrowed when $\beta>1$ and kept unchanged when $\beta=1$.
If $\beta$ is a function of time $\tau$,  the received signal is $f_1(B(\tau))$. Inferred from~\cite{Xia1992On}, if $\beta(\tau)$ is not constant, $f_1(B(\tau))$ is a non-bandlimited signal. Thus, a bandlimited signal becomes non-bandlimited when it is transmitted through Schwarzschild spacetime. Clearly,  the higher the varying rate of $\beta(\tau)$ is, the higher the non-bandlimited-ness is. The contribution to the non-bandlimited-ness consists of gravitationally-modified Doppler effect and  gravitational effect. If the varying rate of $\beta_1(\tau)$ (or $\beta_2(\tau)$) is large, the non-bandlimited-ness contributed by $\beta_1(\tau)$ (or  $\beta_2(\tau)$ ) is significant; if it is small,  the non-bandlimited-ness contributed by $\beta_1(\tau)$ (or  $\beta_2(\tau)$ ) may be negligible.

\subsection{Variation of a signal for straight trajectory} {\label{3.1}} 

\begin{figure}[h!]
\centering
\includegraphics[scale=0.7]{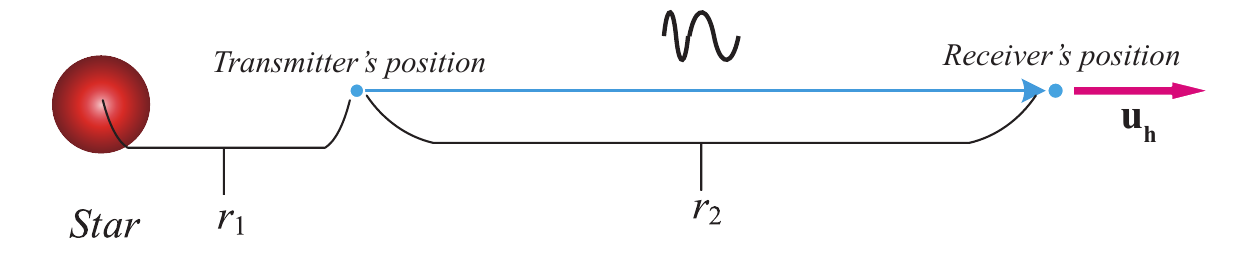}
\caption{A spaceship moving away from a star and communicating with a transmitter at a fixed position.}
\label{flying_away}
\end{figure}

In this section, we consider one of the simplest cases 
 that a receiver  is moving away from a star with  $dr_2/dT\triangleq u_h$ being constant, and a transmitter's position $r_1$, the receiver's position $r_2$ and the center of the star are at the same line as shown in Fig. \ref{flying_away}. Suppose when the receiver's time $\tau=0$, $T$ as a function of $\tau$ is zero and the receiver begins to receive a signal from the transmitter at position $r_0$. At Newtonian time $T(\tau)$ when the receiver's time is $\tau$, 
\begin{equation}{\label{eq38}}
\begin{small}
r_2=r_0+u_hT(\tau).
\end{small}
\end{equation}
From  (\ref{eq13}) and (\ref{eq38}),
\begin{equation}{\label{eq39}}
\begin{small}
\begin{aligned}
u_g
&=(1-\alpha/(r_0+u_hT(\tau)))^{-1}dr_2/dT\\
&=(1-\alpha/(r_0+u_hT(\tau)))^{-1}u_h.
\end{aligned}
\end{small}
\end{equation}
Newtonian time $T(\tau)$ as a function of $\tau$ can be obtained from  (\ref{eq18}), (\ref{eq38}) and (\ref{eq39}):
\begin{equation}{\label{eq40}}
\begin{small}
\begin{aligned}
d\tau=&(1-u_g^2/c^2)^{\frac{1}{2}}(1-\alpha/r_2)^{\frac{1}{2}}dT\\
=&[1-\alpha/(r_0+u_hT(\tau))-(1-\alpha/(r_0+u_hT(\tau)))^{-1}\frac{u_h^2}{c^2}]^{\frac{1}{2}}dT,
\end{aligned}
\end{small}
\end{equation}
where the relation between $\tau$ and $T$ depends on $u_h$ and $r_0$, and as long as $u_h$ and $r_0$ are provided, $T$ as a function of $\tau$ is fixed. $\cos\psi=1$ in  (\ref{eq19}) . Therefore, from (\ref{eq21}), (\ref{eq24}), (\ref{eq38}), (\ref{eq39}) and (\ref{eq40}), GMDFSP and GFSP are, respectively,
\begin{equation}{\label{eq42}}
\begin{small}
\begin{aligned}
\beta_1(\tau)=&(1-u_g^2/c^2)^{-\frac{1}{2}}(1-u_g/c) \\
=&[1-(1-\alpha/(r_0+u_hT(\tau)))^{-2}u_h^2/c^2]^{-\frac{1}{2}}\\
&\times[1-(1-\alpha/(r_0+u_hT(\tau)))^{-1}u_h/c]\\
=&\bigg(\frac{1-\alpha/(r_0+u_hT(\tau))-u_h/c}{1-\alpha/(r_0+u_hT(\tau))+u_h/c}\bigg)^{\frac{1}{2}},\\
\beta_2(\tau)=&\bigg(\frac{1-\alpha/r_{1}}{1-\alpha/(r_{0}+u_hT(\tau))}\bigg)^{\frac{1}{2}}.
\end{aligned}
\end{small}
\end{equation}
 Since $T(\tau)$ just depends on the coefficients $u_h$ and $r_0$, the coefficients that FSP $\beta(\tau)=\beta_1(\tau)\beta_2(\tau)$ depends on are also  $u_h$ and $r_0$. 

\begin{figure*}[t!]
\centering
\includegraphics[scale=0.6]{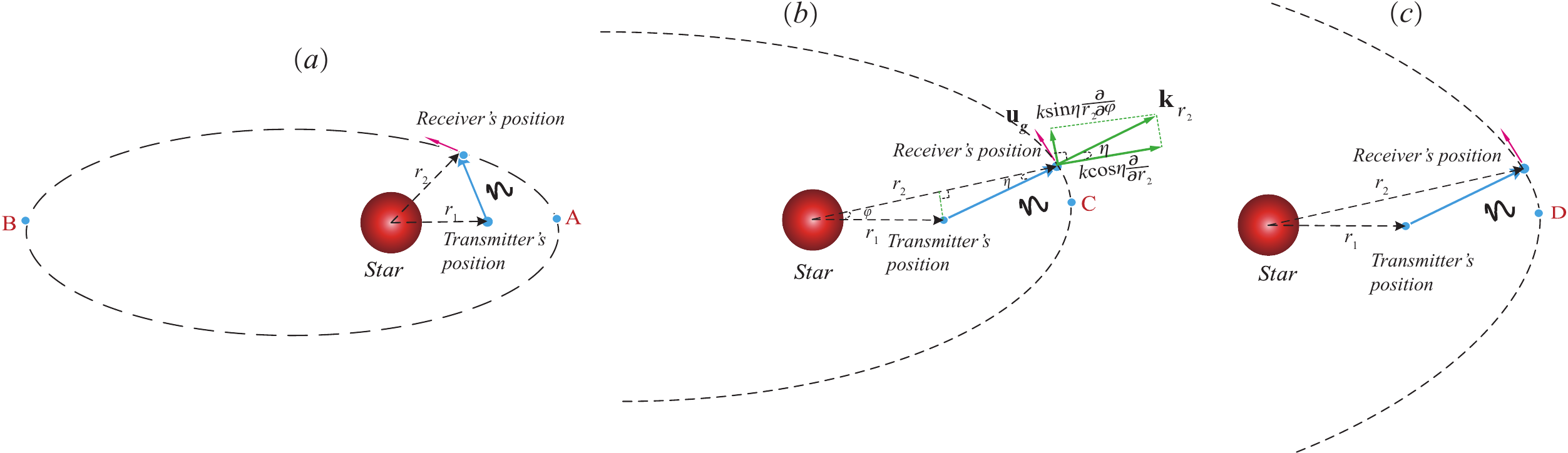}
\caption{A spaceship moving around a star with $(a)$ an elliptical, $(b)$ a parabolic or $(c)$ a  hyperbolic trajectory and communicating with a transmitter at a fixed position.}
\label{conic}
\end{figure*}
No speed is bigger than the speed of light, i.e., $u_h<c$. If $u_h\ll c$ , then
\begin{equation}{\label{eq43}}
\begin{small}
\begin{aligned}
&\bigg(\frac{1-\alpha/(r_0+u_hT(\tau))-u_h/c}{1-\alpha/(r_0+u_hT(\tau))+u_h/c}\bigg)^{\frac{1}{2}}\\
\approx& 1-\left(1-\alpha/(r_0+u_hT(\tau))\right)^{-1}u_h/c,
\end{aligned}
\end{small}
\end{equation}
where $\left(1-\alpha/(r_0+u_hT(\tau))\right)^{-1}$ is not far away from 1 in general, and $\beta(\tau)$  can be written as
\begin{equation}{\label{eq44}}
\begin{small}
\begin{aligned}
\beta(\tau)\approx&\bigg(1-\frac{u_h/c}{1-\alpha/(r_0+u_hT(\tau))}\bigg)\left(\frac{1-\alpha/r_{1}}{1-\alpha/(r_{0}+u_hT(\tau))}\right)^{\frac{1}{2}},  
\end{aligned}
\end{small}
\end{equation}
and
\begin{equation}{\label{eq45}}
\begin{small}
d\tau\approx (1-\alpha/(r_0+u_hT(\tau)))^{\frac{1}{2}}dT.
\end{small}
\end{equation}

Since the studied stars are not black hole, $\alpha<r_0$,  as  mentioned in Section \ref{section2}. If $\alpha\ll r_0$, i.e., the star is far away from being a black hole, then from  (\ref{eq44}),
\begin{equation}{\label{eq46}}
\begin{small}
\begin{aligned}
\beta(\tau)\approx& (1-\alpha/r_{1})^{\frac{1}{2}}\bigg(1-\frac{u_h}{c}-\frac{\alpha u_h/c}{r_0+u_hT(\tau)}\bigg)\bigg(1+\frac{\alpha/2}{r_{0}+u_hT(\tau)}\bigg)\\
\approx&(1-\alpha/r_{1})^{\frac{1}{2}}\left[1-\frac{u_h}{c}+\left(1-\frac{3u_h}{c}\right)\frac{\alpha/2}{r_0+u_hT(\tau)}\right],
\end{aligned}
\end{small}
\end{equation}
and from (\ref{eq45}),
\begin{equation}{\label{eq47}}
\begin{small}
d\tau\approx(1-\alpha/[2(r_0+u_hT(\tau)])dT.
\end{small}
\end{equation}
Since $T(0)=0$, 
\begin{equation}{\label{eq48}}
\begin{small}
\tau\approx T(\tau)-\frac{\alpha}{2u_h}\ln(1+u_hT(\tau)/r_0).
\end{small}
\end{equation}

Furthermore, if $u_hT(\tau)\ll r_0$, i.e., the flying distance of the spaceship during the time of communications is relatively short, then from (\ref{eq46}),
\begin{equation}{\label{eq49}}
\begin{small}
\begin{aligned}
\beta(\tau)\approx& (1-\alpha/r_{1})^{\frac{1}{2}}\bigg[1-\frac{u_h}{c}+\frac{1}{2}\bigg(1-\frac{3u_h}{c}\bigg)\frac{\alpha/r_0}{1+u_hT(\tau)/r_0}\bigg]\\
\approx& (1-\alpha/r_{1})^{\frac{1}{2}}\bigg[1-\frac{u_h}{c}+\frac{\alpha}{2r_0}\bigg(1-\frac{3u_h}{c}\bigg)\bigg(1-\frac{u_hT(\tau)}{r_0}\bigg)\bigg]\\
=&C_0+C_1T(\tau),
\end{aligned}
\end{small}
\end{equation}
where 
\begin{equation}{\label{eq50}}
\begin{small}
\begin{aligned}
C_0&\triangleq\left(1-\alpha/r_{1}\right)^{\frac{1}{2}}\left[1-\frac{u_h}{c}+\frac{\alpha}{2r_0}\left(1-\frac{3u_h}{c}\right)\right],\\
C_1&\triangleq-\frac{u_h\alpha}{2r_0^2}\left(1-\alpha/r_{1}\right)^{\frac{1}{2}}\left(1-\frac{3u_h}{c}\right),
\end{aligned}
\end{small}
\end{equation}
and from (\ref{eq48}),
\begin{equation}{\label{eq51}}
\begin{small}
\begin{aligned}
\tau&\approx \left(1-\alpha/(2r_0)\right)T(\tau)=C_2T(\tau),
\end{aligned}
\end{small}
\end{equation}
where $C_2\triangleq1-\alpha/(2r_0)$. Consequently, after a signal $f_1(t_1)$ is transmitted at $r_1$ to the spaceship at $r_2(\tau)$, the received signal by the spaceship is
\begin{equation}{\label{eq53}}
\begin{small}
f_2(\tau)\approx f_1\left(C_0\tau+C_3\tau^2\right),
\end{small}
\end{equation}
where $C_3\triangleq C_1/(2C_2)$. It can be seen from~\cite{Xia1992On} that if $f_1(t_1)$ is bandlimited, then $f_2(\tau)$ is not bandlimited anymore. Is $f_1(t_1)$  bandlimited in fractional Fourier domain~\cite{Almeida1994The,Tao2007Spectral}? To answer this question, firstly consider a single-frequency signal 
\begin{equation*}
\begin{small}
f_1(t_1)=e^{j\omega t_1},
\end{small}
\end{equation*}
the received signal by the spaceship is 
\begin{equation*}
\begin{small}
f_2(\tau)\approx e^{j\omega (C_0\tau+C_3\tau^2)},
\end{small}
\end{equation*}
which is  a chirp signal with initial frequncy $C_0\omega$ and frequency rate $C_3\omega$. Therefore,      $f_2(\tau)$ is bandlimited in the fractional Fourier domain with frequency rate  $C_3\omega$.
However, if $f_1(t_1)$ is not single-frequency, then $f_2(\tau)$  is  non-bandlimited in any fractional Fourier domain. To see this, let us consider the following signal of two single frequencies: 
\begin{equation*}
\begin{small}
\begin{aligned}
f_1(t_1)&=e^{j\omega_1 t_1}+e^{j\omega_2 t_2}=p_1(t_1)+q_1(t_1),
\end{aligned}
\end{small}
\end{equation*}
with $\omega_1\neq\omega_2$. Then, the received signal by the spaceship is
\begin{equation*}
\begin{small}
\begin{aligned}
f_2(\tau)&\approx e^{j\omega_1 (C_0\tau+C_3\tau^2)}+e^{j\omega_2 (C_0\tau+C_3\tau^2)}=p_2(\tau)+q_2(\tau).
\end{aligned}
\end{small}
\end{equation*}
Notably, $p_2(\tau)$ and $q_2(\tau)$ are both chirp signals but with different frequency rates; therefore it is impossible to find a common fractional domain where both fractional Fourier transforms of $p_2(\tau)$ and $q_2(\tau)$ are bandlimited~\cite{Xia1996On}. As a consequence, $f_2(\tau)$ is not bandlimited in fractional Fourier domain. This applies to any non-single-frequency signal.


\subsection{Variation of a signal for conic trajectories}{\label{3.2}} 
In general, when moving around a star with no extra force, a spaceship may follow a conic trajectory. Therefore, it is worthwhile calculating the change of a signal when the spaceship follows such a trajectory. There are three different kinds of trajectories: ellipse, parabola and hyperbola as shown in Fig. \ref{conic}. Whether it is elliptical, parabolic or hyperbolic depends on the mechanical energy of the spaceship which is equal to the sum of gravitational potential energy ($<0$) and kinetic energy ($>0$). If the mechanical energy is smaller than 0, then the trajectory is elliptical. If the mechanical energy is equal to 0, then the trajectory is parabolic. If the mechanical energy is bigger than 0, then the trajectory is hyperbolic.

Consider the transmitter's position is just a fixed position close to a star and a spaceship with the receiver aboard is moving around the star with a conic trajectory, and its radius relative to the center of the star is~\cite{Carroll2014An} 
\begin{equation}{\label{eq55}}
\begin{small}
r_2=p/(1+e\cos\varphi),
\end{small}
\end{equation}
and
\begin{equation}{\label{eq56}}
\begin{small}
 h=r_2^{2}d\varphi/dT,
 \end{small}
\end{equation}
where $p=h^2/k^2$ and $e=[1+2h^2E/(k^4m)]^{\frac{1}{2}}$ is the eccentricity of the trajectory (ellipse, $0<e<1$; parabola, $e=1$; hyperbola, $e>1$), and $m$ is the mass of the spaceship. The mechanical energy $E$ and the angular momentum $h$ are constants based on mechanical energy conservation and angular momentum conservation, respectively, $T$ is Newtonian time, and $k^2=GM$ where $M$ is the mass of the star. If the semi-major axis $a$ of its orbit is known, then
\begin{equation}{\label{eq57}}
\begin{small}
p=\left\{
\begin{aligned}
&a(1-e^2)&&\rm for&&0<e<1,\\
&2a &&\rm for&&e=1,\\
&a(e^2-1)&&\rm for&& e>1.
\end{aligned}
\right.
\end{small}
\end{equation}

Combining (\ref{eq55}) and (\ref{eq56}), the angular velocity in Newtonian time is 
\begin{equation}{\label{eq58}}
\begin{small}
\frac{d\varphi}{dT}=\frac{h}{p^2}\left(1+e\cos\varphi\right)^2.
\end{small}
\end{equation}
The radial velocity of the spaceship relative to the center of the star  in Newtonian time is 
\begin{equation}{\label{eq59}}
\begin{small}
\begin{aligned}
\frac{d r_2}{dT}&=\frac{pe\sin\varphi}{(1+e\cos\varphi)^{2}} \frac{d\varphi}{dT}=\frac{h}{p}e\sin\varphi.
\end{aligned}
\end{small}
\end{equation}
Since $\theta=\pi/2$, according to (\ref{eq12}), (\ref{eq13}),  (\ref{eq55}), (\ref{eq58}) and (\ref{eq59}),
\begin{equation}{\label{newneweq57}}
\begin{small}
{\bf u_g}=\frac{h}{p}(1-\alpha/r_2)^{-\frac{1}{2}}[e\sin\varphi\partial/\partial r_2+r_2^{-2}p\partial/\partial\varphi]
\end{small}
\end{equation}
and, 
\begin{equation}{\label{eq63}}
\begin{small}
\begin{aligned}
u_g=\frac{h}{p}(1-\alpha/r_2)^{-\frac{1}{2}}\big[(1-\alpha/r_2)^{-1}e^2\sin^2\varphi+(1+e\cos\varphi)^2\big]^{\frac{1}{2}}.
\end{aligned}
\end{small}
\end{equation}
Then, according to  (\ref{eq18}), (\ref{eq55}), (\ref{eq58}) and (\ref{eq63}),
\begin{equation}{\label{eq64}}
\begin{small}
\begin{aligned}
d\tau
=&\bigg\{1-\alpha/r_2-\frac{h^2}{c^2p^2}\big[(1-\alpha/r_2)^{-1}e^2\sin^2\varphi\\
&+(1+e\cos\varphi)^2\big]\bigg\}^{\frac{1}{2}}\frac{p^2}{h}\left(1+e\cos\varphi\right)^{-2}d\varphi.
\end{aligned}
\end{small}
\end{equation}
Suppose the transmitted signal approximately propagates along straight lines. Then, since $\theta=\pi/2$, in (\ref{eq9}), let  
\begin{equation}{\label{neweq57}}
\begin{small}
{\bf k}_{r_2}=k\cos\eta\partial/\partial r_2 +k\sin\eta r_2^{-1}\partial/\partial \varphi.
\end{small}
\end{equation}
where $k$ is a parameter and $\eta$ is the intersection angle between ${\bf k}_{r_2}$ and the radial direction at $r_2$. From Fig. \ref{conic} (b), it is not hard to see,
\begin{equation}{\label{neweq58}}
\begin{small}
\begin{aligned}
\cos\eta=&\frac{r_2-r_1\cos\varphi}{(r_1^2+r_2^2-2r_1r_2\cos\varphi)^{\frac{1}{2}}},\\
\sin\eta=&\frac{r_1\sin\varphi}{(r_1^2+r_2^2-2r_1r_2\cos\varphi)^{\frac{1}{2}}}.
\end{aligned}
\end{small}
\end{equation}
According to (\ref{eq6}) with ${\bf k}={\bf k}_{r_2}$ and $\omega=\omega_2$, (\ref{neweq57}) and (\ref{neweq58}),
\begin{equation}{\label{neweq59}}
\begin{small}
k=\frac{\omega_2}{c}\frac{(r_1^2+r_2^2-2r_1r_2\cos\varphi)^{\frac{1}{2}}}{[(1-\alpha/r_2)^{-1}(r_2-r_1\cos\varphi)^2+r_1^2\sin^2\varphi]^{\frac{1}{2}}}
\end{small}
\end{equation}
Then, according to (\ref{newneweq57}), (\ref{neweq57}),  (\ref{neweq58}) and (\ref{neweq59}), $\cos\psi$ in  (\ref{eq19}) is
\begin{equation}{\label{neweq63}}
\begin{small}
\begin{aligned}
\cos\psi=&\frac{{\bf g}({\bf k}_{r_2}, {\bf u_g})}{\sqrt{{\bf g}({\bf k}_{r_2}, {\bf k}_{r_2}){\bf g}({\bf u_g}, {\bf u_g})}}=\frac{{\bf g}({\bf k}_{r_2}, {\bf u_g})}{u_g\omega_2/c},
\end{aligned}
\end{small}
\end{equation}
where
\begin{equation}
\begin{small}
\begin{aligned}
{\bf g}({\bf k}_{r_2}, {\bf u_g})=&\frac{h\omega_2}{pc}(1-\alpha/r_2)^{-\frac{3}{2}}\sin\varphi[r_2e-r_1(1+e\cos\varphi)\alpha/r_2\\
&+r_1][(1-\alpha/r_2)^{-1}(r_2-r_1\cos\varphi)^2+r_1^2\sin^2\varphi]^{-\frac{1}{2}}.
\end{aligned}
\end{small}
\end{equation}
As long as the initial value of $\varphi$ and the values of $h$ and $p$ (or $a$ and $e$) are given, $\varphi$ as a function of $\tau$ can be obtained from  (\ref{eq55}) and (\ref{eq64}). Next, it can be seen from  (\ref{eq55}), (\ref{eq63}) and (\ref{neweq63}) that $r_2$, $u_g$ and $\cos\psi$ are functions of $\varphi$; therefore, they are also functions of $\tau$. Neglecting the hindrance by the star on signals, if the spaceship does not use any extra force, according to  (\ref{eq20}),  FSP as a function of $\tau$ is 
\begin{equation}{\label{eq65}}
\begin{small}
\begin{aligned}
\beta(\tau)
&=(1-u_g^2/c^2)^{-\frac{1}{2}}\left(1-\frac{u_g}{c}\cos{\psi}\right)\left(\frac{1-\alpha/r_{1}}{1-\alpha/r_{2}}\right)^{\frac{1}{2}},
\end{aligned}
\end{small}
\end{equation}
and 
\begin{equation}{\label{eq66}}
\begin{small}
\begin{aligned}
\beta_{1}(\tau)=&(1-u_g^2/c^2)^{-\frac{1}{2}}\left(1-\frac{u_g}{c}\cos{\psi}\right),\\
\beta_{2}(\tau)=&\left(\frac{1-\alpha/r_{1}}{1-\alpha/r_{2}}\right)^{\frac{1}{2}},
\end{aligned}
\end{small}
\end{equation}
are the GMDFSP and GFSP, respectively.

\begin{figure*}[t!]
\centering
\includegraphics[scale=0.7]{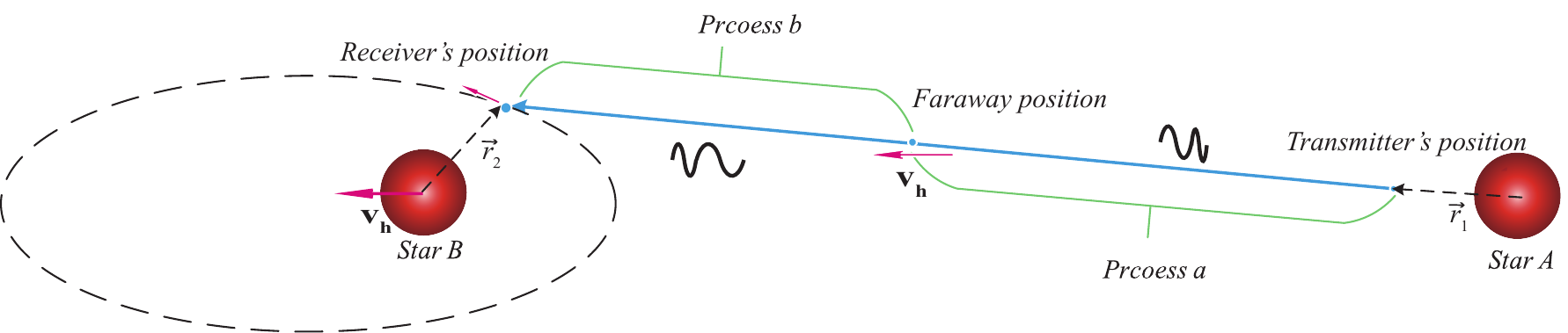}
\caption{A signal propagating from star system A to another faraway star system B.}
\label{two_stars}
\end{figure*}
\subsection{Variation of a signal for interstellar communications} 
Mankind may have chances to receive signals from aliens in another faraway star system who may find it not so hard to send an electromagnetic wave containing information to solar system. Besides, mankind may be able to settle on many other star systems in the future, so communications between different star systems may be frequent. Therefore, It is worth calculating the change of a signal between different star systems.

Consider a signal is transmitted from a star system A to another faraway star system B and the relative velocity ${\bf v_h}$ between the two stars is fixed with a magnitude $v_h$  in Newtonian time $T$ along a fixed direction as shown in Fig. \ref{two_stars}. Since the two stars are far away from each other, the gravitation of one star can be neglected if the signal is much closer to the other star. When the signal is near the transmitter, the gravitational  effect is almost caused by star A. When the signal is near the receiver, the gravitational  effect is almost caused by star B. However, when the signal is neither near star A nor star B, the gravitation can be neglected. Therefore, the whole process can be separated into two parts: a) the signal is propagating from the transmitter's position near star A to a position which is far away from star A and star B (called the faraway position in the following) and is moving with the same velocity ${\bf v_h}$ as star B; b) the signal is propagating from the faraway position to the receiver's position near star B. In the process a), it is the same as that considered in Section~\ref{3.1} as shown in~Fig.\ref{flying_away}. Set  the parameter of the frequency shift as $\beta_a$. Since the faraway position $r_\infty\rightarrow \infty$ and the faraway position is moving away from transmitter position with velocity ${\bf v_h}$, according to  (\ref{eq20}), if the position of transmitter is fixed, then 
\begin{equation}{\label{eq67}}
\begin{small}
\begin{aligned}
\beta_a
=&(1-v_h^2/c^2)^{-\frac{1}{2}}(1-v_h/c)(1-\alpha_1/r_1)^{\frac{1}{2}},
\end{aligned}
\end{small}
\end{equation}
where $\alpha_1=2GM_1/c^2$, $M_1$ is the mass of star A, $r_1$ is the radial distance of the transmitter relative to the center of star A.

In the process b),  it is similar to that in Section~\ref{3.2} as shown in Fig. \ref{conic} (a). Set the parameter of the frequency shift as $\beta_b$.   Suppose the receiver at position $r_2$ follows an elliptical trajectory, according to  (\ref{eq55}) 
\begin{equation}{\label{eq68}}
\begin{small}
\begin{aligned}
r_2&=p/(1+e\cos\varphi),
\end{aligned}
\end{small}
\end{equation}
where $0<e<1$. Let ${\bf u_g}$ be the spatial velocity of the receiver in Schwarzschild spacetime, according to  (\ref{eq63}),
\begin{equation}{\label{eq69}}
\begin{small}
\begin{aligned}
u_g=&\frac{h}{p}(1-\alpha_2/r_2)^{-\frac{1}{2}}\big[(1-\alpha_2/r_2)^{-1}e^2\sin^2\varphi+(1+e\cos\varphi)^2\big]^{\frac{1}{2}}.
\end{aligned}
\end{small}
\end{equation}
Substituting $r_1$ in  (\ref{neweq63}) for $r_\infty$,
\begin{equation}{\label{eq70}}
\begin{small}
\begin{aligned}
\cos\psi=&\sin\varphi[1-(1-\alpha/r_2)^{-1}\alpha e\cos\varphi/r_2][(1-\alpha/r_2)^{-1}\cos^2\varphi\\
&+\sin^2\varphi]^{\frac{1}{2}}[(1-\alpha/r_2)^{-1}e^2\sin^2\varphi+(1+e\cos\varphi)^2]^{-\frac{1}{2}},
\end{aligned}
\end{small}
\end{equation}
and according to (\ref{eq64}),
\begin{equation}{\label{eq71}}
\begin{small}
\begin{aligned}
d\tau=&\bigg\{1-\alpha_2/r_2-\frac{h^2}{c^2p^2}\big[(1-\alpha_2/r_2)^{-1}e^2\sin^2\varphi\\
&+(1+e\cos\varphi)^2\big]\bigg\}^{\frac{1}{2}}
\frac{p^2}{h}\left(1+e\cos\varphi\right)^{-2}d\varphi,
\end{aligned}
\end{small}
\end{equation}
where $\tau$ is the time of the receiver and $\varphi$ as a function of $\tau$ can be solved. It can be seen from  (\ref{eq68}), (\ref{eq69}) and (\ref{eq70}) that $r_2$, $u_g$ and $\cos\psi$ are functions of $\varphi$, i.e., functions of $\tau$. Therefore,  according to  (\ref{eq20}), $\beta_b$ as a function of $\tau$ is
\begin{equation}{\label{eq72}}
\begin{small}
\begin{aligned}
\beta_b(\tau)
&=(1-u_g^2/c^2)^{-\frac{1}{2}}\left(1-\frac{u_g}{c}\cos{\psi}\right)(1-\alpha_2/r_{2})^{-\frac{1}{2}}.\\
\end{aligned}
\end{small}
\end{equation}
Combining processes a) and b), the FSP of the whole process is 
\begin{equation}{\label{eq73}}
\begin{small}
\begin{aligned}
\beta(\tau)=&\beta_a\beta_b(\tau)\\
=&(1-v_h^2/c^2)^{-\frac{1}{2}}(1-u_g^2/c^2)^{-\frac{1}{2}}(1-v_h/c)\left(1-\frac{u_g}{c}\cos{\psi}\right)\\
&\times\left(\frac{1-\alpha_1/r_1}{1-\alpha_2/r_{2}}\right)^{\frac{1}{2}},\\
\end{aligned}
\end{small}
\end{equation}
and 
\begin{equation}{\label{eq74}}
\begin{small}
\begin{aligned}
\beta_1(\tau)=&(1-v_h^2/c^2)^{-\frac{1}{2}}(1-u_g^2/c^2)^{-\frac{1}{2}}(1-v_h/c)\left(1-\frac{u_g}{c}\cos{\psi}\right),\\
\beta_2(\tau)=&\left(\frac{1-\alpha_1/r_1}{1-\alpha_2/r_{2}}\right)^{\frac{1}{2}},
\end{aligned}
\end{small}
\end{equation}
are the GMDFSP and GFSP, respectively.

\begin{figure*}[!h]
\centering
\includegraphics[scale=0.7]{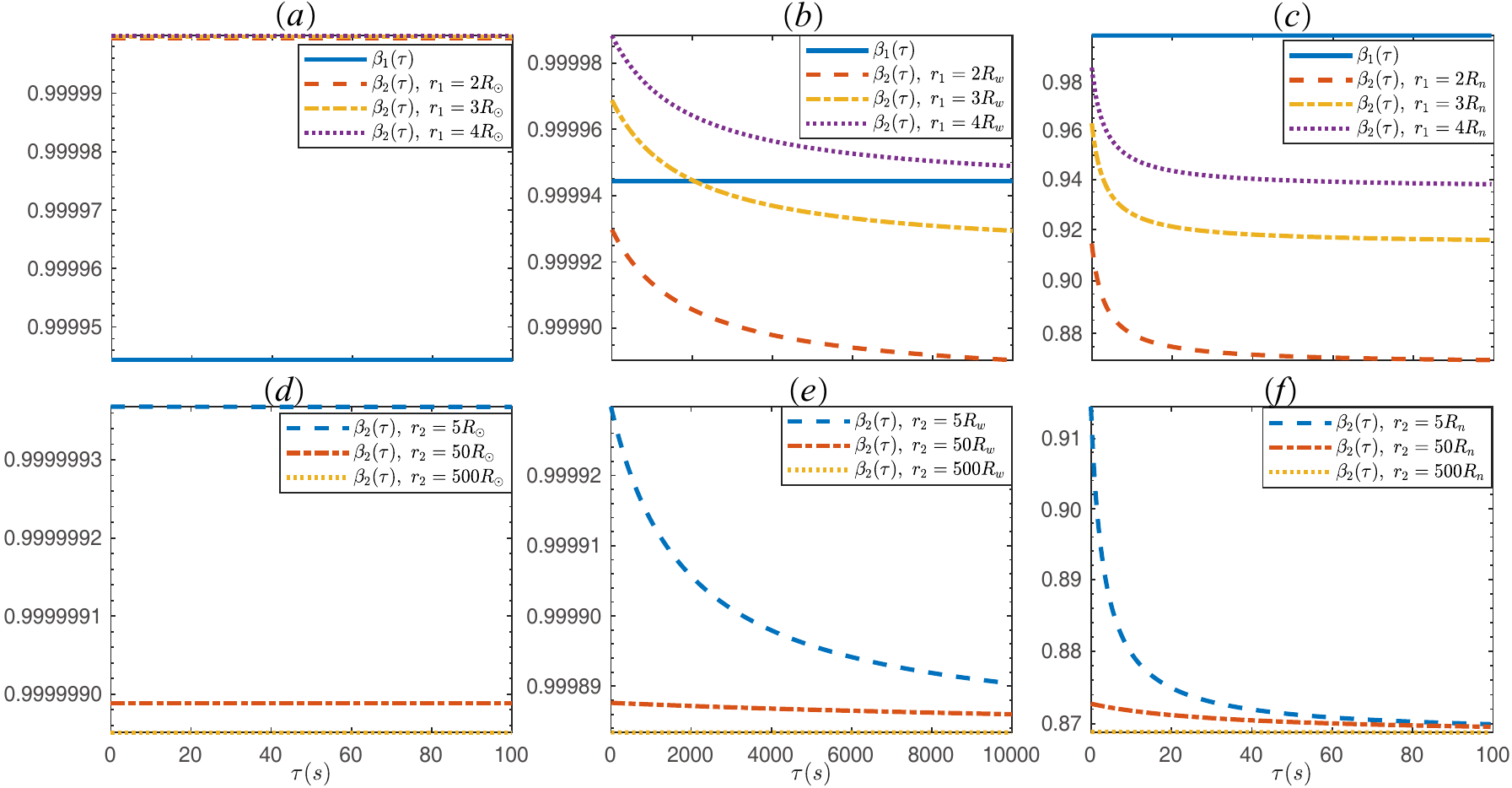}
\caption{The frequency shift parameter from gravitational effect (GFSP) and gravitationally-modified Doppler effect (GMDFSP) vs. time when a spaceship moves away from $(a)$, $(d)$ sun, $(b)$, $(e)$ white dwarf or $(c)$, $(f)$ neutron star for  different positions of a transmitter and  for different initial positions of a receiver, respectively.}
\label{swn1}
\end{figure*}
\subsection{Variation of a signal transmitted from a moving position to a moving spaceship}{\label{3.4}}
In the above studies, a transmitter is at a fixed position. The above results can be easily extended to the case when the transmitter is in motion as follows. 

A moving transmitter has its own 4-velocity and its frequency is just the projection value along its 4-velocity.  Suppose the frequency of a signal that the moving transmitter at $r_1$ transmits is $\omega_1'$. Then, the relation between the frequencies of a static observer at $r_1$ and the moving transmitter at $r_1$ can be obtained by using (19) with $r_2$ replaced by $r_1$:  
\begin{equation}{\label{neweq72}}
\begin{small}
\omega_1=\omega'_1(1-u^2_{1g}/c^2)^{\frac{1}{2}}\left(1-\frac{u_{1g}}{c}\cos\psi_1\right)^{-1},
\end{small}
\end{equation}
where $\cos\psi_1={\bf g}({\bf k}_{r_1}, {\bf u_{1g}})/\sqrt{{\bf g}({\bf k}_{r_1}, {\bf k}_{r_1}){\bf g}({\bf u_{1g}}, {\bf u_{1g}})}$, ${\bf k}_{r_1}$is the spatial wave vector at $r_1$, and ${\bf u_{1g}}$ is the spatial velocity of the moving transmitter in Schwarzschild spacetime and $u_{1g}$ is the magnitude. Then, by using (19) again without modification, i.e., the relation between the frequencies of a moving observer at $r_2$ and a static observer at $r_1$, the frequency of a moving receiver at $r_2$ is
\begin{equation}{\label{neweq73}}
\begin{small}
\omega'_2=\omega_1(1-u^2_{2g}/c^2)^{-\frac{1}{2}}\left(1-\frac{u_{2g}}{c}\cos\psi_2\right)\left(\frac{1-\alpha/r_{1}}{1-\alpha/r_{2}}\right)^{\frac{1}{2}},
\end{small}
\end{equation}
where $\cos\psi_2={\bf g}({\bf k}_{r_2}, {\bf u_{2g}})/\sqrt{{\bf g}({\bf k}_{r_2}, {\bf k}_{r_2}){\bf g}({\bf u_{2g}}, {\bf u_{2g}})}$, ${\bf k}_{r_2}$is the spatial wave vector at $r_2$, and ${\bf u_{2g}}$ is the spatial velocity of the moving receiver in Shwarzschild spacetime and $u_{2g}$ is the magnitude. Then,
\begin{equation}{\label{neweq74}}
\begin{small}
\omega'_2=\omega'_1\bigg(\frac{c^2-u^2_{1g}}{c^2-u^2_{2g}}\bigg)^{\frac{1}{2}}\bigg(\frac{c-u_{2g}\cos\psi_2}{c-u_{1g}\cos\psi_1}\bigg)\left(\frac{1-\alpha/r_{1}}{1-\alpha/r_{2}}\right)^{\frac{1}{2}}.
\end{small}
\end{equation}

Equation (\ref{neweq74}) is extended from (\ref{eq19}). Similar to the three scenarios above, the specific parameters in (\ref{neweq74}) are related to the trajectories of a transmitter and a receiver. 

\section{Numerical examples}{\label{section4}}
In this section, we present some numerical examples to illustrate what we have studied above. 
The studied stars  are sun-like star, white dwarf and neutron star. The mass of the sun $M_\odot$ is approximately $1.9891\times 10^{30}kg$ and its radius $R_\odot$ is about $6.955\times 10^8m$~\cite{Woolfson2000The}. 
The mass of the white dwarf is set as $M_\odot$ and its radius as $0.9\%R_\odot$ which is nearly the radius of the earth.  
The mass of the neutron star $M_n$ is set as $2.01M_\odot$ and its radius $R_n$ as $1.71\times 10^{-5}R_\odot$ or $12km$. All these settings are according to \cite{Shipman1979Masses, Kepler2007White,Ozel2012On, Chamel2013On,antoniadis2013massive,Hasensel2007Neutron,steiner2013neutron,zhao2015properties}.

\subsection{Spaceship moving away from a star  }{\label{section4.1}}
In this section, consider transmitting a signal at a fixed position to a spaceship  moving away from the sun, a white dwarf or a neutron star as shown in Fig. \ref{flying_away}. In all examples $u_h=dr_2/dT$  is set as the third cosmic velocity $16.7km/s$.

\begin{figure*}[!h]
\centering
\includegraphics[scale=0.72]{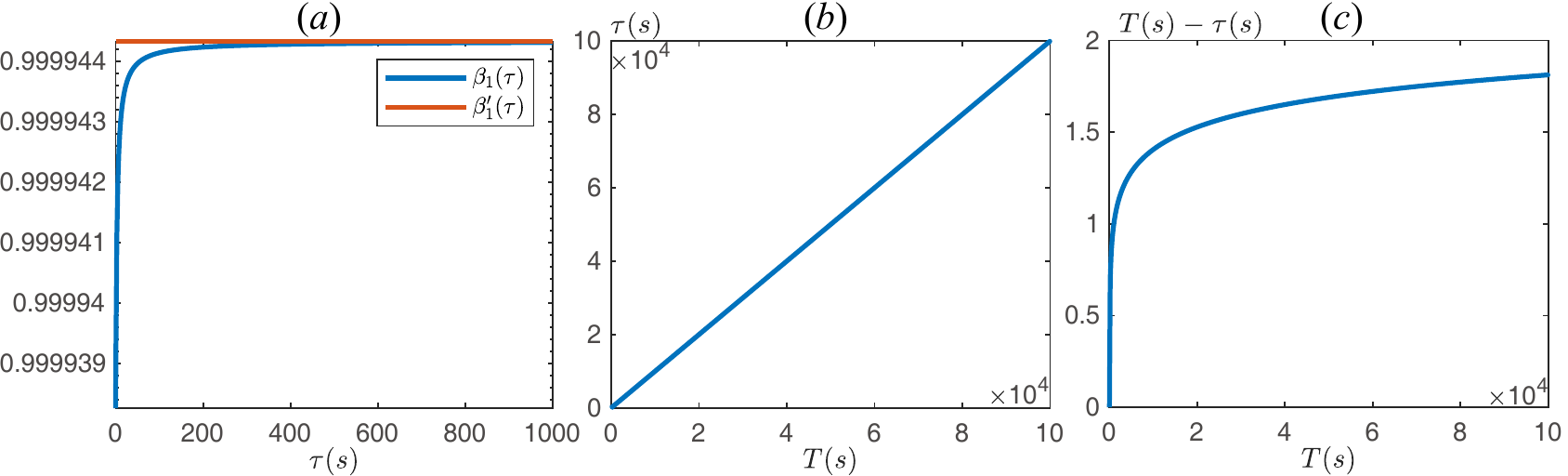}
\caption{$(a)$ The difference between  Doppler effect $\beta'_1(\tau)$ (special relativity version) and gravitationally-modified Doppler effect $\beta_1(\tau)$ as a spaceship moves away from a neutron star with a fixed Newtonian velocity; $(b)$ the time $\tau$ of the moving spaceship vs. Newtonian time $T$; $(c)$ the difference between $\tau$ and $T$ vs. $T$. The initial position of the receiver is $5R_n$.  }
\label{swn2}
\end{figure*} 
\begin{figure*}[!h]
 \centering
     \includegraphics[scale=0.7]{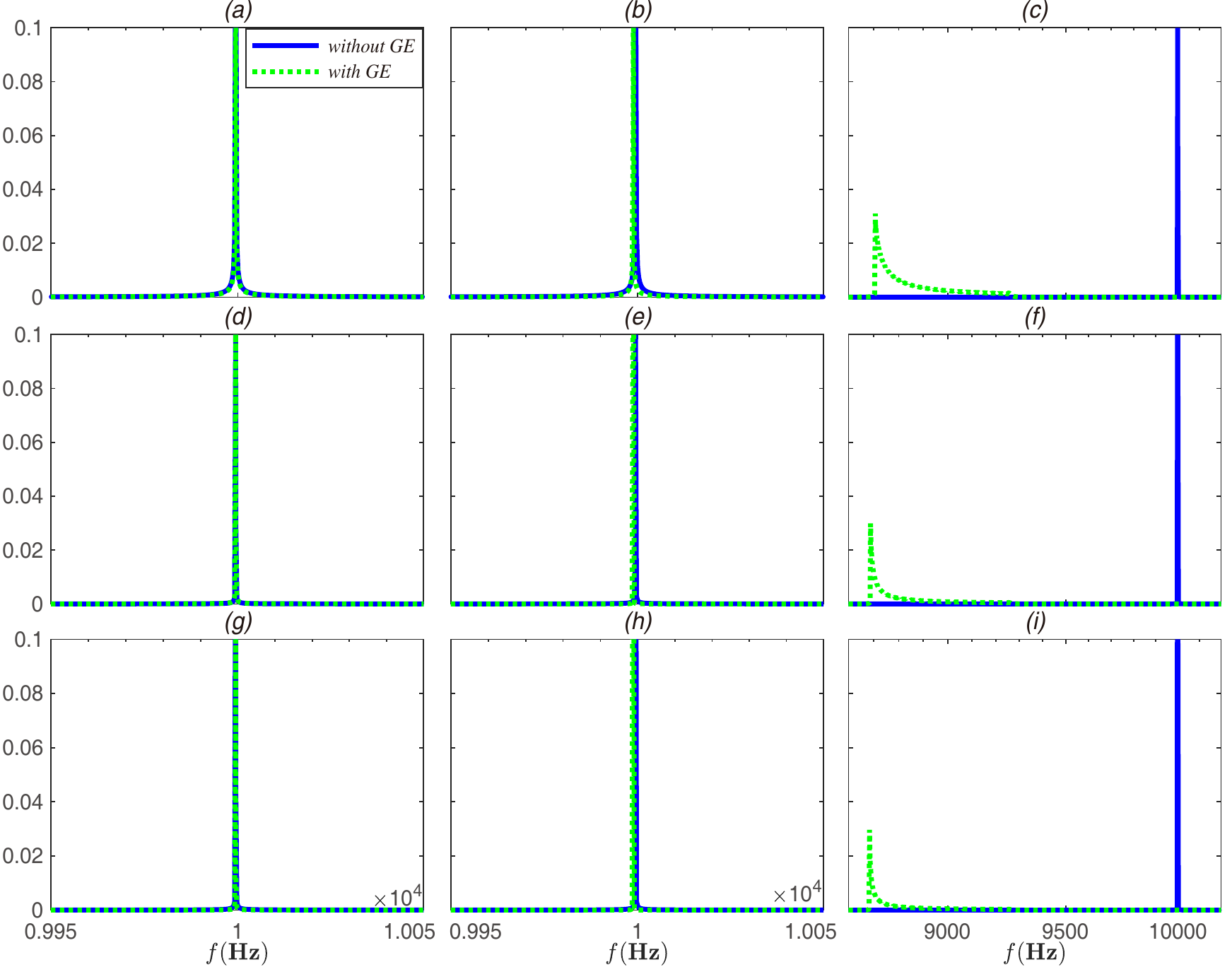}
\caption{The FFTs of a $10^4$ Hz frequency signal  when it is transmitted in $(a)$, $(d)$, $(g)$ the solar, $(b)$, $(e)$, $(h)$ white dwarf or $(c)$, $(f)$, $(i)$ neutron star systems with or without gravitational effects during  $\tau$=50s, 150s and 250s, respectively. The transmitters' positions are $2R_{\odot}$, $2R_w$ and $2R_n$, respectively. The initial positions of the receivers are $4R_{\odot}$, $4R_w$ and $4R_n$, respectively. The sampling rate is 2.5 times of the frequency. }
\label{swn3}
\end{figure*}  

In Figs.~\ref{swn1} $(a)$, $(b)$, $(c)$, the initial distances of the receiver  to the centers of the sun, the white dwarf and the neutron star, when it begins to receive signals, are $5R_\odot$, $5R_w$ and $5R_n$, respectively. Compared to gravitationally-modified Doppler effect , gravitational effect is negligible in the solar system since GFSP $\beta_2(\tau)$ is much close to 1 in all three cases. As $\beta_1(\tau)$ and $\beta_2(\tau)$ are both nearly  constant, the non-bandlimited-ness due to gravitationally-modified Doppler effect and gravitational effect is negligible in the solar system.
In contrast, when the transmitter and the receiver  are both not far from the white dwarf, gravitational effect is comparable to gravitationally-modified Doppler effect and the non-bandlimited-ness contributed by gravitational effect is noticeable. 
 When the transmitter is close to the neutron star, compared to the gravitational effect, gravitationally-modified Doppler effect can be neglected, and 
  the frequency rate contributed by $\beta_2(\tau)$ is huge, 
  which causes a high non-bandlimited-ness. 
In Figs.~\ref{swn1} $(d)$, $(e)$, $(f)$, whether the communications is in the solar system, the white dwarf system or the neutron star system, when the receiver is faraway, $\beta_2(\tau)$ is nearly constant (see the curves of $r_2=500R_\odot$, $500R_w$, $500R_n$). This is because when the receiver is far away from them, the gravitation on the receiver becomes negligible. As a result, the frequency  rate is very small and the non-bandlimited-ness  is less noticeable.

Fig.~\ref{swn2}  is to show the difference of gravitationally-modified Doppler effect $\beta_1(\tau)$ and Doppler effect (special relativity version) $\beta_1'(\tau)$, and the difference of the time $\tau$ of a moving receiver and Newtonian time $T$. Since the gravitation in neutron star system is much higher than that in other star systems,  these differences are much clearer in a neutron star system. Therefore, a neutron star system is chosen to show these differences. In Fig.~\ref{swn2} $(a)$, it shows that gravitationally-modified Doppler effect $\beta_1(\tau)$ is more than Doppler effect $\beta'_1(\tau)$ since $\beta'_1(\tau)$ is closer to 1, but the difference gradually  tends to be zero as the receiver is far away from the neutron star. Although the Fig.~\ref{swn2} $(b)$ shows that $\tau$ is approximately linear to $T$, the difference between $\tau$ and $T$ in Fig.~\ref{swn2} $(c)$ increases drastically in the beginning and then tends to be linear to $T$. 

\begin{figure*}[!t]
\centering
\includegraphics[scale=0.7]{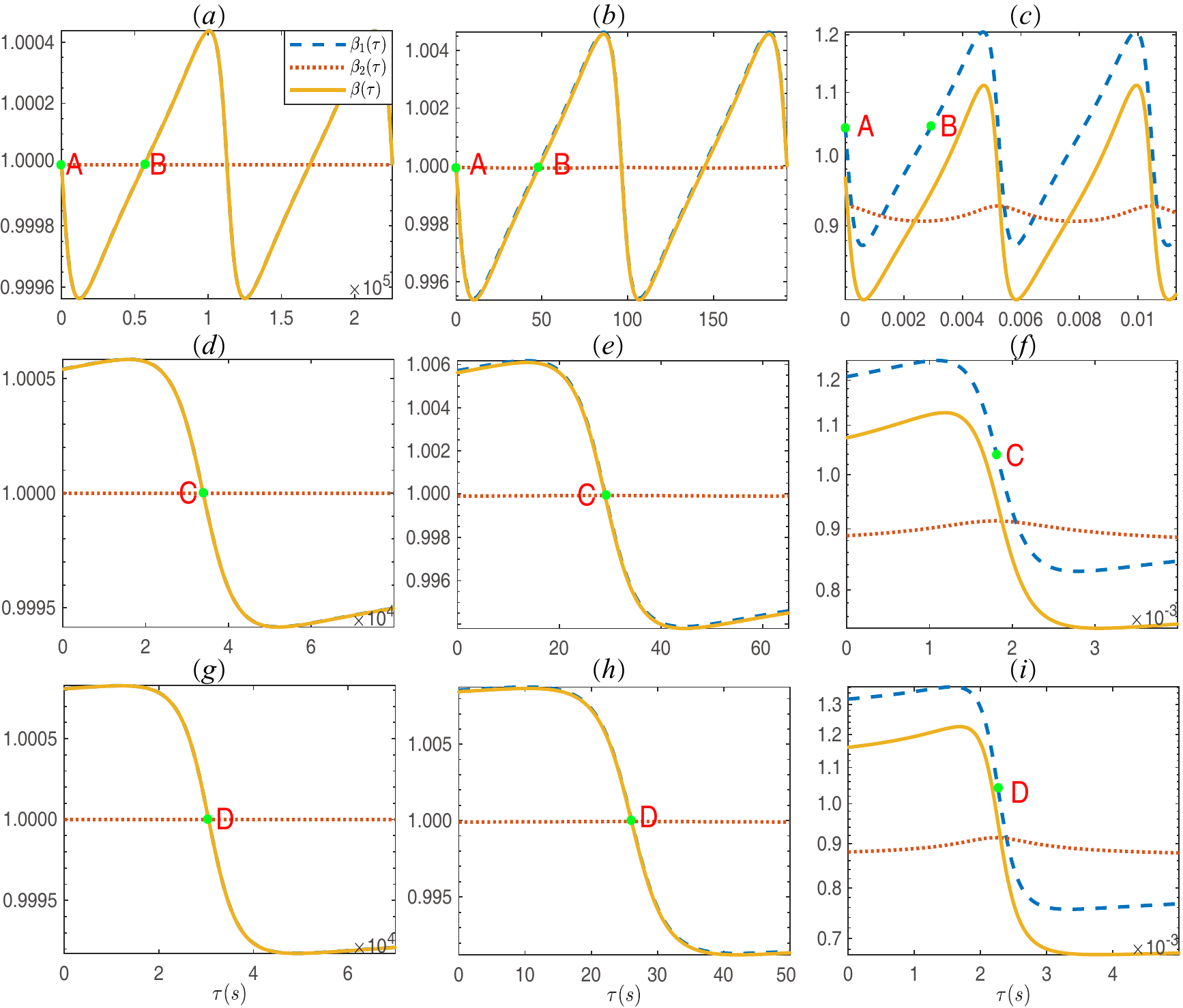}
\caption{The frequency shift parameter from gravitationally-modified Doppler effect (GMDFSP), gravitational effect (GFSP) and  the whole frequency shift parameter (FSP) vs. time when a spaceship moves around $(a)$, $(d)$, $(g)$ sun, $(b)$, $(e)$, $(h)$ white dwarf, or $(c)$, $(f)$, $(i)$ neutron star for different conic trajectories, i.e., ellipse ($e=0.2$), parabola ($e=1$) and hyperbola ($e=2$), and $a=5R_\odot$, $5R_w$ and $5R_n$, respectively. }
\label{swn4}
\end{figure*}
\begin{figure*}[!h]
\centering
\includegraphics[scale=0.68]{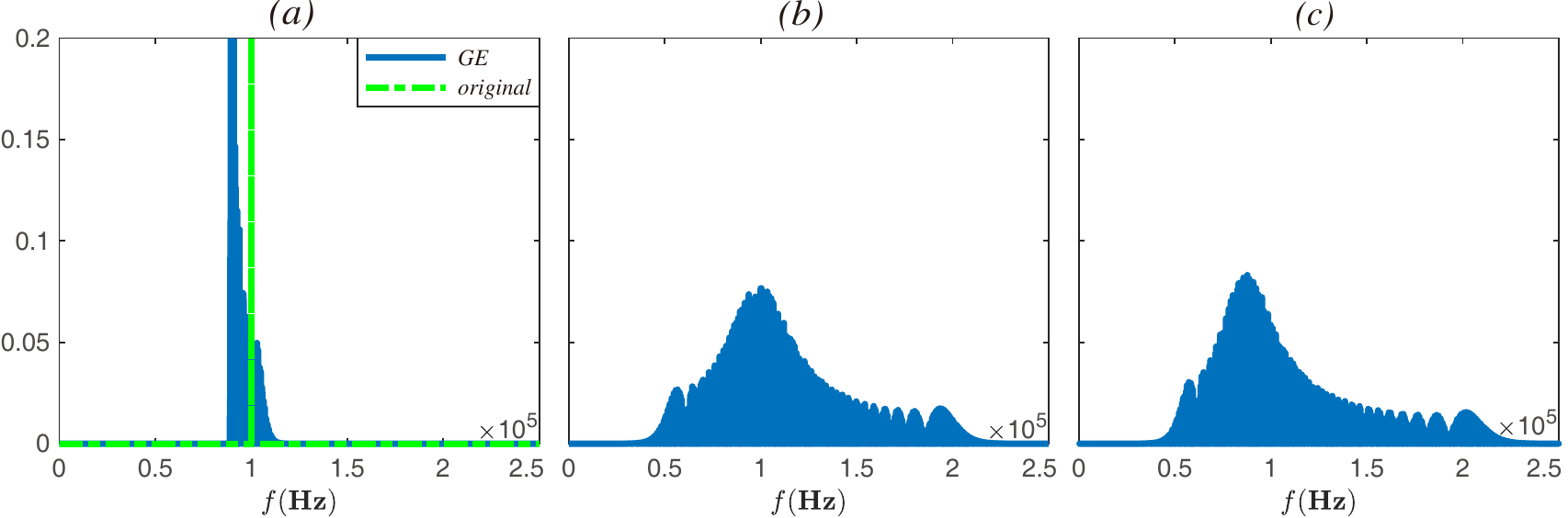}
\caption{The FFTs of a $10^5$Hz signal $(a)$ with gravitational effect only $(\beta_2(\tau))$, $(b)$ with gravitationally-modified Doppler effect only $(\beta_2(\tau))$ and $(c)$ with whole effect $(\beta(\tau))$, when it is transmitted in the neutron star system from a position at $2R_n$ to another position in an elliptic trajectory with $e=0.0167$ and semi-major axis $a=5R_n$. The length of time is $10^4$ periods about 56.54s and the sampling rate is 2.5 times of the frequency.}
\label{swn5}
\end{figure*}

In Fig.~\ref{swn3}, to show the non-bandlimited-ness of a signal in these star systems, the FFTs of a special bandlimited signal---a single-frequency complex exponential signal transmitted in  the solar system, the white dwarf system or the neutron star system during different time periods  are calculated. In the solar system,  the FFTs with and without gravitational effect are nearly the same during the same time periods. The bands with and without gravitational effect are both  very narrow centering around centric frequency $10^4$Hz. In the white dwarf system, the FFTs are close to each other, but the FFT with gravitational effect has a relative translation to the left. The bands with and without gravitational effect are both very narrow while centering around a little different frequencies.  In the neutron star system, when the  period of time is small, the FFT with gravitational effect spans over a broad frequency band and peaks at much smaller frequency, compared with the band without gravitational effect. However, when the period of time is large, the band with gravitational effect is narrowed. 

\begin{figure*}[ht!]
\centering
\includegraphics[scale=0.7]{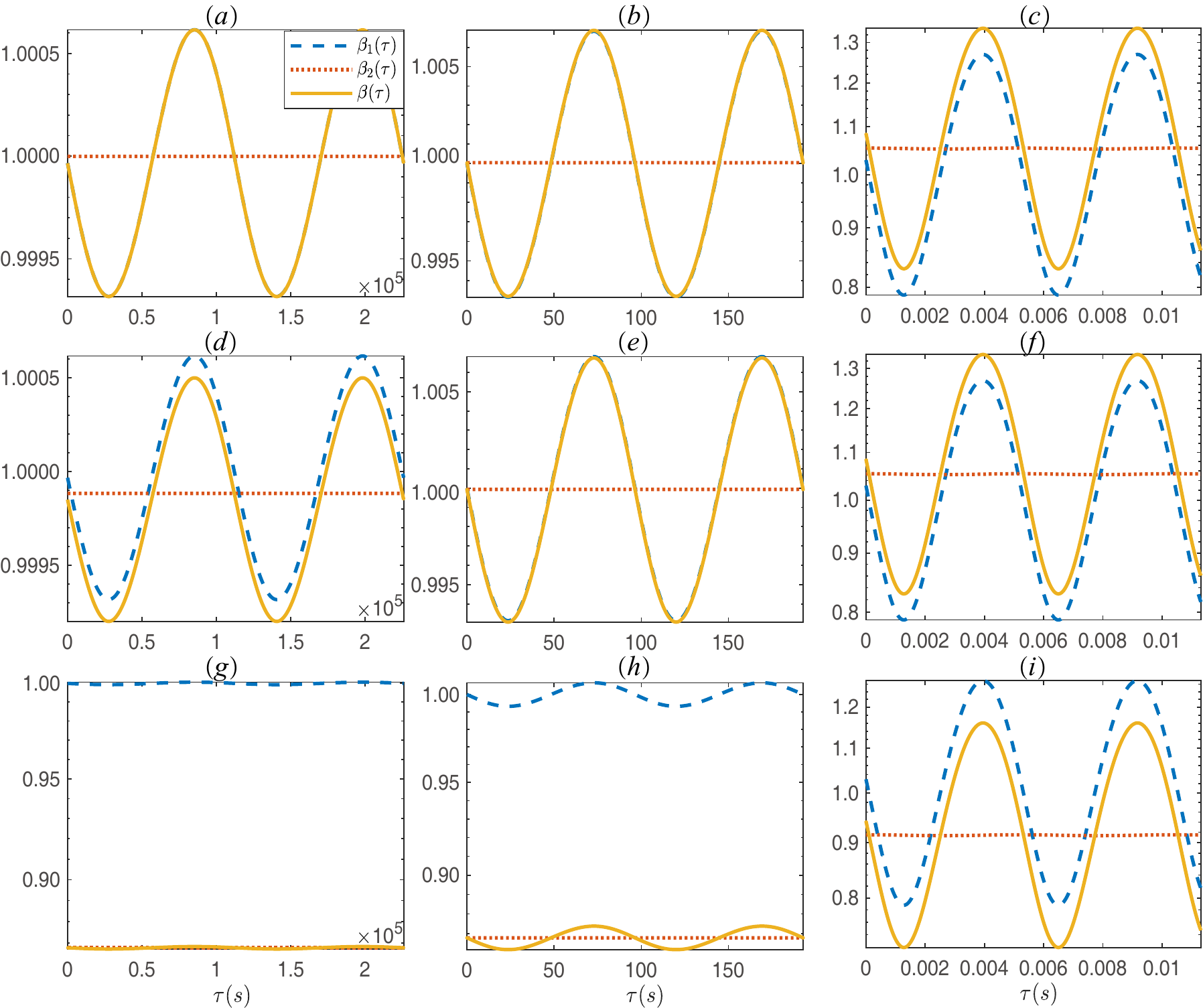}
\caption{The frequency shift parameter from gravitationally-modified Doppler effect (GMDFSP), gravitational effect (GFSP) and  the whole frequency shift parameter (FSP) vs. time for a signal received at a position  with elliptical trajectory ($e=0.0167$) in $(a)$, $(d)$, $(g)$ a sun-like star, $(b)$, $(e)$, $(h)$ a white dwarf or $(c)$, $(f)$, $(i)$ a neutron star system from a fixed position in  another sun-like star,  white dwarf and  neutron star system, respectively. }
\label{swn6}
\end{figure*} 
\subsection{Spaceship moving around a star with  conic trajectories}{\label{section4.2}
In this section, we consider transmitting a signal at a fixed position to a spaceship moving around the sun, a white dwarf or a neutron star with elliptical, parabolic or hyperbolic trajectory as shown in Fig. \ref{conic}.

In Figs. \ref{swn4} $(a), (d), (g)$, compared to gravitationally-modified Doppler effect in all three cases, gravitational effect and the non-bandlimited-ness contributed by gravitational effect are both negligible in the solar system. In Figs.~\ref{swn4} $(b)$, $(e)$, $(h)$, though small, the differences of $\beta(\tau)$ and $\beta_1(\tau)$ can be noticed in all trajectories around the white dwarf. Therefore, gravitational effect should be taken into account in the white dwarf system. However, the non-bandlimited-ness contributed by gravitational effect is negligible in all trajectories because of little variation of $\beta_2(\tau)$.
In Figs.~\ref{swn4} $(c)$, $(f)$, $(i)$, when the spaceship moves around the neutron star, the gravitational effect is significant.  The variation of $\beta_2(\tau)$ is noticeable, especially in the elliptical trajectory, and the non-bandlimited-ness contributed by gravitational effect cannot be ignored. However, the variation of $\beta_1(\tau)$ is more significant. Therefore, the non-bandlimited-ness contributed by gravitationally-modified Doppler effect is more significant.

Whether the star is the sun, the white dwarf or the neutron star, the non-bandlimited-ness is mainly  contributed by gravitationally-modified Doppler effect. Furthermore, in the elliptical trajectories, when the spaceship is around the position that is closest or farthest to the transmitter (see position A and position B in Fig.~\ref{conic} and  Fig.~\ref{swn4}), the frequency rate derived from $\beta_1(\tau)$ is high, which leads to a high non-bandlimited-ness. For parabolic and hyperbolic trajectories, when the spaceship is around the closest position to the transmitter (see position C and position D in Fig.~\ref{conic} and  Fig.~\ref{swn4}), the frequency rates are also high while they are nearly zero when the spaceship is far away from the transmitter. The levels of non-bandlimited-ness in the solar system, the white dwarf system and the neutron star system are different. Obviously, the level of non-bandlimited-ness is the highest in the neutron star system, followed by the white dwarf system. In the solar system, the non-bandlimited-ness can be neglected.

Since gravitational effect in the solar system or the white dwarf system is not significant and the variation of $\beta_2(\tau)$ is more significant in the elliptical trajectory, only the spectrum  of a  signal transmitted to a spaceship following an elliptical trajectory in the neutron star system is analyzed.  In Fig.~\ref{swn5},  when only taking gravitational effect $\beta_2(\tau)$ into account, the single-frequency complex exponential signal expands over a noticeable frequency band. This confirms that gravitational effect has a contribution to the non-bandlimited-ness. When only taking gravitationally-modified Doppler effect $\beta_1(\tau)$ into account, the signal expands over even more wider frequency band. The whole effect $\beta(\tau)$  has the similar effect as gravitationally-modified Doppler effect, which confirms that the non-bandlimited-ness is mainly determined by gravitationally-modified Doppler effect.

\subsection{Communications between different star systems}
In this section, gravitational effect on a signal is calculated when it is transmitted from a sun-like star, a white dwarf or a neutron star system to another  sun-like star, white dwarf or neutron star system as shown in Fig. \ref{two_stars}. The receiver moves with an elliptical trajectory. The mass and the radius of the sun-like stars are the same as the sun. Suppose the communications lasts from a few seconds to a few minutes. $v_h$ in  (\ref{eq73}) is set as $10^4km/s$.

In Figs.~\ref{swn6} $(a)$, $(d)$, $(g)$,  a signal sent by intelligent species in another system comes to the solar system during a relatively short period of time. 
If the signal  comes from a sun-like system, gravitational effect can be neglected. 
If the signal comes from the white dwarf system, gravitational effect is comparable to gravitationally-modified Doppler effect. 
 If the signal comes from the neutron star system, gravitational effect is significant 
 and gravitationally-modified Doppler shift is relatively small.  However, whatever the signal source system is, the FSP $\beta (\tau)$ is nearly constant during a relative small period of time. 
As a result, the frequency rate is nearly zero and the non-bandlimited-ness is negligible.
  
In Figs.~\ref{swn6} $(b)$, $(e)$, $(h)$, a signal from a star system is transmitted to a white dwarf system. On one hand, if the source star system is a sun-like star system or a white dwarf system, gravitational effect can be simply neglected. 
On the other hand, if the source star is a neutron star, gravitational effect is significant, compared to gravitationally-modified Doppler effect. 
 However,  $\beta_2(\tau)$ is nearly constant and the non-bandlimited-ness contributed by gravitationally-modified Doppler effect is dominant. Therefore, neither gravitational effect nor gravitationally-modified Doppler effect can be ignore, and gravitational effect provides large frequency shift and gravitationally-modified Doppler effect provides high non-bandlimited-ness. In any case, the frequency rate is noticeable and the non-bandlimited-ness is significant.

In Figs.~\ref{swn6} $(c)$, $(f)$, $(i)$, a neutron star receives a signal from a star system. On one hand, for all three cases, 
gravitationally-modified Doppler effect  
and the non-bandlimited-ness contributed by gravitationally-modified Doppler effect are both significant. 
 On the other hand, for all the three cases,  the differences of $\beta_2(\tau)$ and 1 are all large and gravitational effect is significant, 
but $\beta_2(\tau)$ is nearly constant, compared to $\beta_1(\tau)$.  Therefore, the non-bandlimited-ness is mainly contributed by gravitationally-modified Doppler effect. 

\section{Conclusion}{\label{section5}}
In this paper, the variation of a signal in Schwarzschild spacetime was studied and a general equation for frequency shift parameter (FSP), which can be divided into gravitational FSP and gravitationally-modified Doppler FSP, was presented. In addition, rates of time of a transmitter and a receiver may be different.  Based on the equation for FSP, FSP as a function of the time of a receiver was calculated for scenarios a), b), and c) as described in Abstract. In scenario a), compared to gravitationally-modified Doppler effect, gravitational effect is negligible in the solar system, comparable in a white dwarf system, and more significant in a neutron star system, which suggests that for deep space missions in the solar system like Mars exploration, gravitational effect can be neglected. The non-bandlimited-ness is only contributed by gravitational effect and is more significant in a neutron star system. Furthermore, choosing a neutron star as an example,   when a receiver is close to the neutron star, the difference between gravitationally-modified Doppler effect and the special relativity version of Doppler effect, and the difference between time $\tau$ of a moving receiver and Newtonian time $T$ are both significant. In  scenario b), gravitational effect, similar to scenario a), is insignificant in the solar system, less significant in a white dwarf system and significant in a neutron star system. The non-bandlimited-ness is mainly contributed by gravitationally-modified Doppler effect in all star systems, but the non-bandlimited-ness in the neutron star system cannot be neglected. In scenario c), when  communications is between a sun-like star system and another sun-like star system (or a white dwarf system and another white dwarf system), gravitational effect is negligible; when it is between a white dwarf and a sun-like star, gravitational effect is insignificant if a receiver is in a white dwarf system while significant if a receiver is in a sun-like star system; when it involves a neutron star, gravitational effect cannot be ignored and is comparable to gravitationally-modified Doppler effect. In all the cases of scenario c), the non-bandlimited-ness is mainly contributed by gravitationally-modified Doppler effect.

Although in the most of this paper, the position of a transmitter was fixed, as mentioned and studied in Section~\ref{3.4}, all the results can be easily extended to the case when a transmitter moves. Finally, the studies in this paper  may help to better understand the signal models in deep space communications.   
\ifCLASSOPTIONcaptionsoff
  \newpage
\fi

\end{document}